\documentclass{bmcart}

\usepackage{booktabs} 
\usepackage{graphicx}
\usepackage{subfigure}
\usepackage{algorithm}
\usepackage[noend]{algpseudocode}
\usepackage{pifont}
\usepackage{setspace}
\usepackage{color}
\usepackage{amsmath}
\usepackage{amsfonts}
\usepackage{amssymb}
\usepackage{bm}
\usepackage{multirow}
\usepackage{amsthm}
\usepackage{paralist}

\newtheorem{lemma}{Lemma}

\theoremstyle{definition}
\theoremstyle{plain}

\usepackage{color} 
\usepackage{colortbl}
\usepackage{cases}
\definecolor{light-gray}{gray}{0.6}
\definecolor{lavender}{rgb}{0.5,0.5,1.0}

\usepackage{url}

\newcommand{\tb}{\textbf}

\newcommand{\bsw}{\begin{sideways}}
\newcommand{\esw}{\end{sideways}}

\usepackage{xspace}
\newcommand{\HON}{\texttt{BuildHON}\xspace}
\newcommand{\HONp}{\texttt{BuildHON+}\xspace}

\begin{document}

\begin{frontmatter}
\begin{artnotes}
\note[id=n1]{Equal contributor} 
\end{artnotes}
\begin{fmbox}
\dochead{Research}

\title {Efficient Modeling of Higher-Order Dependencies in Networks: From Algorithm to Application for Anomaly Detection}

\author[
    addressref={aff1},
    email={msaebi@nd.edu},
    noteref={n1} ]{\inits{MS}\fnm{Mandana} \snm{Saebi}}
\author[
    addressref={aff1},
    email={jxu5@nd.edu},
        noteref={n1}]{\inits{JX}\fnm{Jian} \snm{Xu}}
\author[
    addressref={aff3},
    email={lance.m.kaplan.civ@mail.mil}
    ]{\inits{LMK}\fnm{Lance M} \snm{Kaplan}}
\author[
    addressref={aff2},
    email={ribeiro@cs.purdue.edu}
    ]{\inits{BR} \fnm{Bruno} \snm{Ribeiro}}
\author[
    addressref={aff1},
    corref={aff1},
    email={nchawla@nd.edu}
    ]{\inits{NVC}\fnm{Nitesh V} \snm{Chawla}}

\address[id=aff1]{%
    \orgname{University of Notre Dame}
    \city{Notre Dame},
    }
\address[id=aff2]{%
    \orgname{Purdue University}
    \city{West Lafayette},
    }
\address[id=aff3]{%
    \orgname{U.S. Army Research Lab}
    }

\end{fmbox}

\ifx
\author{Jian Xu}
\affiliation{%
  \institution{University of Notre Dame}
}
\email{jxu5@nd.edu}

\author{Mandana Saebi}
\affiliation{%
  \institution{University of Notre Dame}
}
\email{mandana.saebi.1@nd.edu}

\author{Bruno Ribeiro}
\affiliation{%
  \institution{Purdue University}
}
\email{ribeiro@cs.purdue.edu}

\author{Lance M. Kaplan}
\affiliation{%
  \institution{U.S. Army Research Lab}
  \postcode{20783}
}
\email{lance.m.kaplan.civ@mail.mil}

\author{Nitesh V. Chawla}
\affiliation{%
  \institution{University of Notre Dame}
}
\email{nchawla@nd.edu}

\renewcommand{\shortauthors}{J. Xu et al.}
\fi

\begin{abstractbox}
\begin{abstract}

Complex systems, represented as dynamic networks, comprise of components that influence each other via direct and/or indirect interactions. Recent research has shown the importance of using Higher-Order Networks (HONs) for modeling and analyzing such complex systems, as the typical Markovian assumption in developing the First Order Network (FON) can be limiting. This higher-order network representation not only creates a more accurate representation of the underlying complex system, but also leads to more accurate network analysis. In this paper, we first present a scalable and accurate model, \HONp,  for higher-order network representation of data derived from a complex system with various orders of dependencies.  Then, we show that this higher-order network representation modeled by \HONp is significantly more accurate in identifying anomalies than FON, demonstrating a need for the higher-order network representation and modeling of complex systems for deriving meaningful conclusions.  

\end{abstract}
\end{abstractbox}

\end{frontmatter}



\section{Introduction}

Networks have become a popular way of representing rich and sparse interactions among the components of a complex system. It is critical for the network to truly represent the inherent phenomena in the complex system to avoid incorrect conclusions.
Conventionally, edges in networks represent the pairwise interactions of the nodes -- assuming the naive Markovian property for node interactions-- resulting in the first-order network representation (FON). However,  the key question is --- {\it{is this accurately representing the underlying phenomena in the complex systems?}} And if the network is not accurately representing the inherent dependencies in the complex system, can we trust the analysis and results stemming from this network? 

Recent research has brought to fore challenges with the FON view, especially its limitations inadequately capturing such higher-order dependencies-- in a complex system. This has led to the development of network representation models that capture such higher-order dependencies, going beyond the traditional pairwise Markovian network representation ~\cite{lambiotte2019networks, xu2016representing}. The Markovian assumption for network modeling of complex system poses several limitations for network analysis, including community detection~\cite{rosvall2014memory,benson2016higher}, node ranking~\cite{scholtes2016higher}, representation learning~\cite{saebi2020honem} and dynamic processes~\cite{scholtes2014causality} in time-varying complex systems.

The higher-order network representation also posits the challenge of scale (computational complexity) to incorporate both variable and higher-orders in the raw data. The scalability of the algorithms is important for enabling efficient network analysis. Our prior work~\cite{xu2016representing} challenges the Markovian assumption for node interactions, and proposes \HON  for extracting higher-order dependencies from sequential data to build the Higher-Order Network (HON) representation. \HON, although accurate, faced the challenge of computational complexity as well as parameter dependency.  In this work, we address the limitations of our prior work by proposing a scalable and parameter-free algorithm, \HONp for accurate extraction of higher-order dependencies from sequential data. Given \HONp, we ask the following research question in this paper as well: 
 {\it{Does incorporating higher-order dependencies improve the performance of existing network-based methods for detecting anomalous signals in the sequential data?}}

To answer the above question, we define anomalies (or change points) as deviations from the norm or expected behavior of a complex system. We note that the anomalies could also be important change points in the behavior of the complex system. The key here is to be able to accurately flag such deviations or events in a complex system. While there exists a wide range of anomaly detection methods on dynamic networks \cite{akoglu2015graph,ranshous2015anomaly}, all of them use the first-order network (FON) to represent the underlying raw data (such as clickstreams, taxi movements, or event sequences), which can lose important higher-order information~\cite{xu2016representing,rosvall2014memory}. As FON is an oversimplification of higher-order dynamics, we hypothesize that anomaly detection algorithms that rely on FONs will miss important changes in the network, thus leaving anomalies undetected. We systematically demonstrate why existing network-based anomaly detection methods can leave certain signals undetected, and propose a higher-order network anomaly detection framework. Consider the following example.

\begin{figure}
\centering
\includegraphics[width=.5\linewidth]{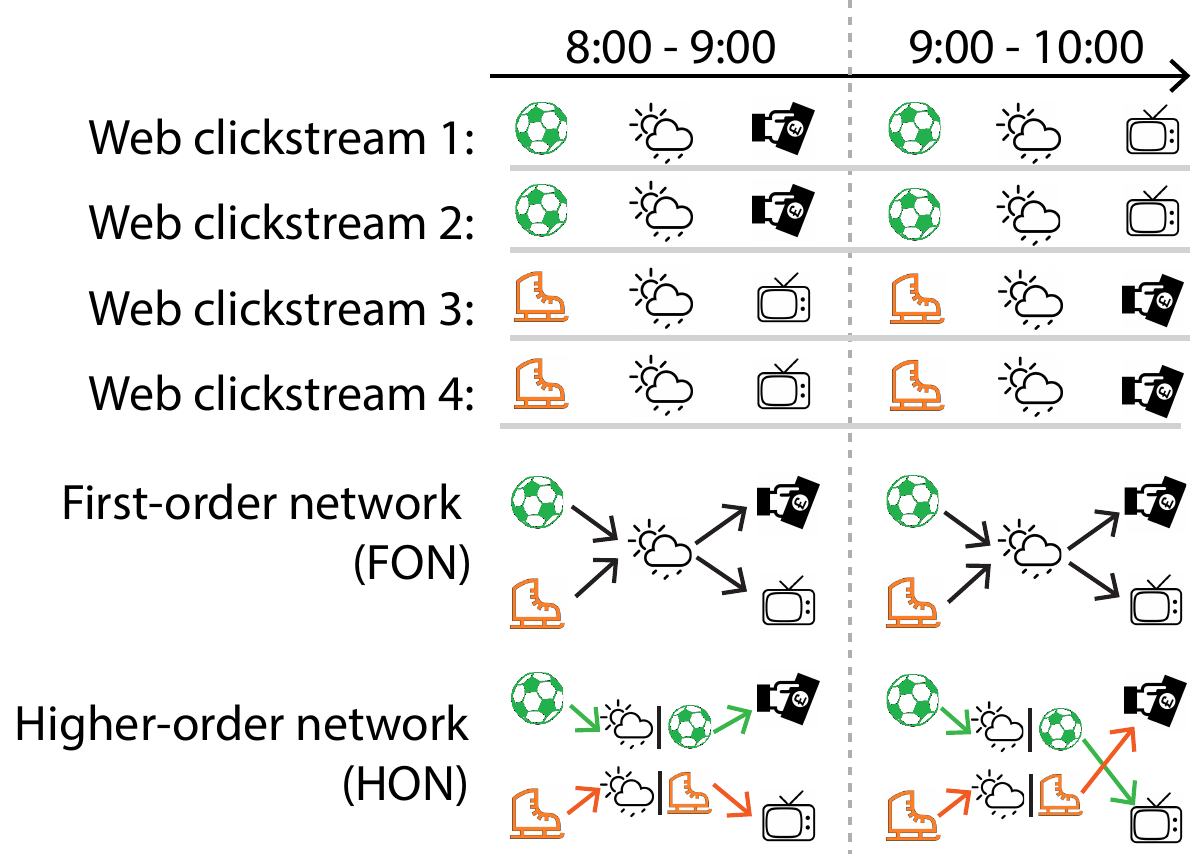}
\caption{\csentence{Higher-order anomalies cannot be detected by network-based anomaly detection methods if FON is used.}}
      \vspace{-.2in}
\label{fig:intro}
\end{figure}

\tb{Example.  }Fig.~\ref{fig:intro} illustrates the challenge of detecting certain types of anomalies, using a minimal example of web clickstreams data (sequences of web page views produced by users) collected by a local media company.
Given the web clickstreams as the input to network-based anomaly detection methods, conventionally, a web traffic network is built for each time window (two one-hour windows illustrated here), with the nodes representing web pages and the edges representing total traffic between web pages.
A change in the network topology indicates an anomaly in web traffic patterns.
According to the original clickstreams, in the first hour, all users coming from the soccer web page to the weather page proceed to the ticket page, and all users coming from the skating page to the weather page go to TV schedules.
But the flow of users is completely flipped in the next hour, possibly the weather forecast has updated with much colder weather which is in favor of winter activities.
However, despite the significant changes in user web viewing patterns, the pairwise traffic between web pages in this example remains the same, thus the FON topology shows no changes.
Therefore, no matter what network-based anomaly detection method is used, if the method relies on FON, the company will not be able to detect such type of anomalies, thus failing to respond (e.g., caching pages for visits, or targeted promotion of pages) to the changes in user behaviors.

\tb{Contributions.}  
We make three main contributions in the paper. 
\begin{itemize}
\item We develop a scalable and parameter-free algorithm for higher-order network representation, \HONp, building on our prior work~\cite{xu2016representing}. We demonstrate the efficiency of \HONp through comprehensive complexity and performance analysis on the global ship movement data, which is known to exhibit dependencies beyond the fifth order. 

\item We showcase the performance of \HONp in the task of network-based anomaly detection on a real-world taxi trajectory data. We explain why the parameter dependency in our prior work can be limiting for efficient network construction and as a result, anomaly detection. 

\item Using a large-scale synthetic taxi movement data with 11 billion taxi movements, we show how multiple existing anomaly detection methods that depend on FON collectively fail to capture anomalous navigation behaviors beyond first-order, and how \HONp can solve the problem.
\end{itemize}

\section{Related Work}
\tb{Higher-order networks.  } Recent research has highlighted the limitations of the conventional network model for representing the sequential and indirect dependencies between the components of complex systems. Multi-layer higher-order models~\cite{kivela2014multilayer,de2016physics}, motif and clique-based higher-order models~\cite{benson2016higher,arenas2008motif,petri2013topological}, and non-Markovian higher-order models~\cite{xu2016representing,scholtes2014causality,rosvall2014memory} try to embed complex patterns that are stemming from the raw data into the network representation. 
Specifically, non-Markovian network models has gained a lot of attraction in many applications including social networks~\cite{karsai2012correlated,wei2015measuring}, human transportation networks~\cite{scholtes2014causality,rosvall2014memory,xu2016representing,matamalas2016assessing}, trade networks~\cite{koher2016infections,lebacher2019exploring}, species flow networks~\cite{saebi2020higher,saebi2020network}, and citation networks~\cite{rosvall2014memory}. Several research studies show how incorporating higher-order dependencies affects various network analysis tasks, including community detection~\cite{rosvall2014memory,benson2016higher}, node ranking~\cite{scholtes2016higher}, representation learning~\cite{saebi2020honem}, and dynamic processes~\cite{scholtes2014causality} in the network. However, from current research studies, it is unclear what is the effect of using a higher-order network model on detecting anomalies in dynamic networks.

\tb{Anomaly detection in dynamic networks.  }
Unlike the task of detecting anomalous nodes and edges in a single static network (such as \cite{akoglu2010oddball}), anomaly detection in dynamic networks \cite{chandola2012anomaly,akoglu2015graph} uses multiple snapshots of networks to represent the interactions of interest (such as interacting molecules \cite{ramanathan2010online}, elements in frames of videos \cite{saligrama2012video}, flow of invasive species \cite{xu2014improving}, etc.), then identifies the time when the network topology shows significant changes, using network distance metrics \cite{shoubridge2002detection,kraetzl2006modality,pincombe2005anomaly}, probability methods \cite{peel2015detecting}, subgraph methods like ~\cite{mongiovi2013netspot} and more.
There are many advantages of using network-based methods for the task of anomaly detection in sequential data. Aside from the availability of several different networks, a graph structure represents the relational nature of the data, which is essential for addressing the anomaly detection problem~\cite{akoglu2015graph}.
Furthermore, the inter-dependencies of the raw data can be captured more efficiently with graph representation. This feature can be further enhanced in the higher-order representation of the network, as done in this work. The importance of higher-order patterns in different network analysis tasks has gained a lot of attention recently~\cite{lambiotte2019networks,scholtes2017network}. However, one of the major challenges is that the graph search space is very large, requiring the anomaly detection methods to be scalable and efficient for large data sets~\cite{akoglu2015graph}.

Moreover, using snapshots of networks may cause the fine-grained time-stamps to be lost. Therefore, the optimal time-stamp is often data-dependent and should be identified empirically through sufficient experiments.

Nevertheless, existing methods on anomaly detection rely on conventional FON; as we will show, certain types of anomalies cannot be detected with any network-based anomaly detection methods if FON is used.
Rather than proposing another approach to identify the anomalous network from a series of networks, our innovation lies in the network construction step, which ensures anomalous signals are preserved in the network in the first place.

\ifx
\tb{Higher-order networks.  } We developed HON in our previous work,~\cite{xu2016representing} as a practical framework for capturing higher-order dependencies in sequential data and representing them as a network. Unlike the conventional FON in which every node represents a single state, a node in the higher-order network can represent the current {\em and} previous states (illustrated in Fig.~\ref{fig:intro}); therefore, the network topology also contains valuable dependency rules in the raw sequential data.
Note that, in  HON representation variable orders of dependencies can appear in the network, since we do not assign a fixed order to the sequential data.  In fixed-order networks \cite{rosvall2014memory}, every node stores memories of the previous $k$ steps; in HON with variable orders \cite{xu2016representing}, higher-order nodes and edges are added only when higher-orders significantly differ from lower orders. 

Nevertheless, Our previously developed algorithm, \HON cannot be adapted directly to network-based anomaly detection due to (1) parameter-dependent results and (2) scalability constraints. \HON requires two parameters that had to be specified experimentally, depending on the data set. Furthermore, it uses an exhaustive search for extracting dependency rules and network construction. In this work, through rigorous optimization and making insightful observations, we have leveraged HON for the anomaly detection task and made it scalable for big data applications.
\fi

\section{Methods}

We first present a scalable and parameter-free approach for constructing HON, namely \HONp. We then show how this new approach enables more accurate anomaly detection (compared to using FON) by incorporating several different network distance measures.
Our previous algorithm, \HON required two parameters that had to be specified experimentally, depending on the data set. Furthermore, it uses an exhaustive search for extracting the dependency rules and constructing the network, which becomes impractical for various network analysis tasks, including anomaly detection. It needs two parameters in addition to the detection threshold: a {\em MaxOrder} parameter which governs how many orders of dependencies the algorithm will consider in HON, and a {\em MinSupport} parameter that discards infrequent observations. These limitations mitigate its applicability to Big Data. 

\subsection{\HONp: Building HON from Big Data}

Here we introduce \HONp, a parameter-free algorithm that constructs HON from big data sets. 
\HONp is a practical approach that preserves higher-order signals in the network representation step ($S_i \rightarrow G_i$) which is essential for anomaly detection. 
The difference between \HON and \HONp is similar to the difference between pruning and early stopping in decision trees. \HON first builds a HON of all orders from first-order to {\em MaxOrder} and then selects branches showing significant higher-order dependencies. \HONp reduces the search space beforehand by checking in each step if increasing the order may produce significant dependencies.
Furthermore, \HON can only discover dependencies up to {\em MaxOrder}. \HONp however, finds the appropriate dependency order hidden in the raw data and is not limited by {\em MaxOrder}. Therefore, the output network resulting from \HONp is a more reliable and accurate representation of the raw data, which is essential for the task of anomaly detection. 

\begin{figure*}
\centering
\includegraphics[scale=0.3]{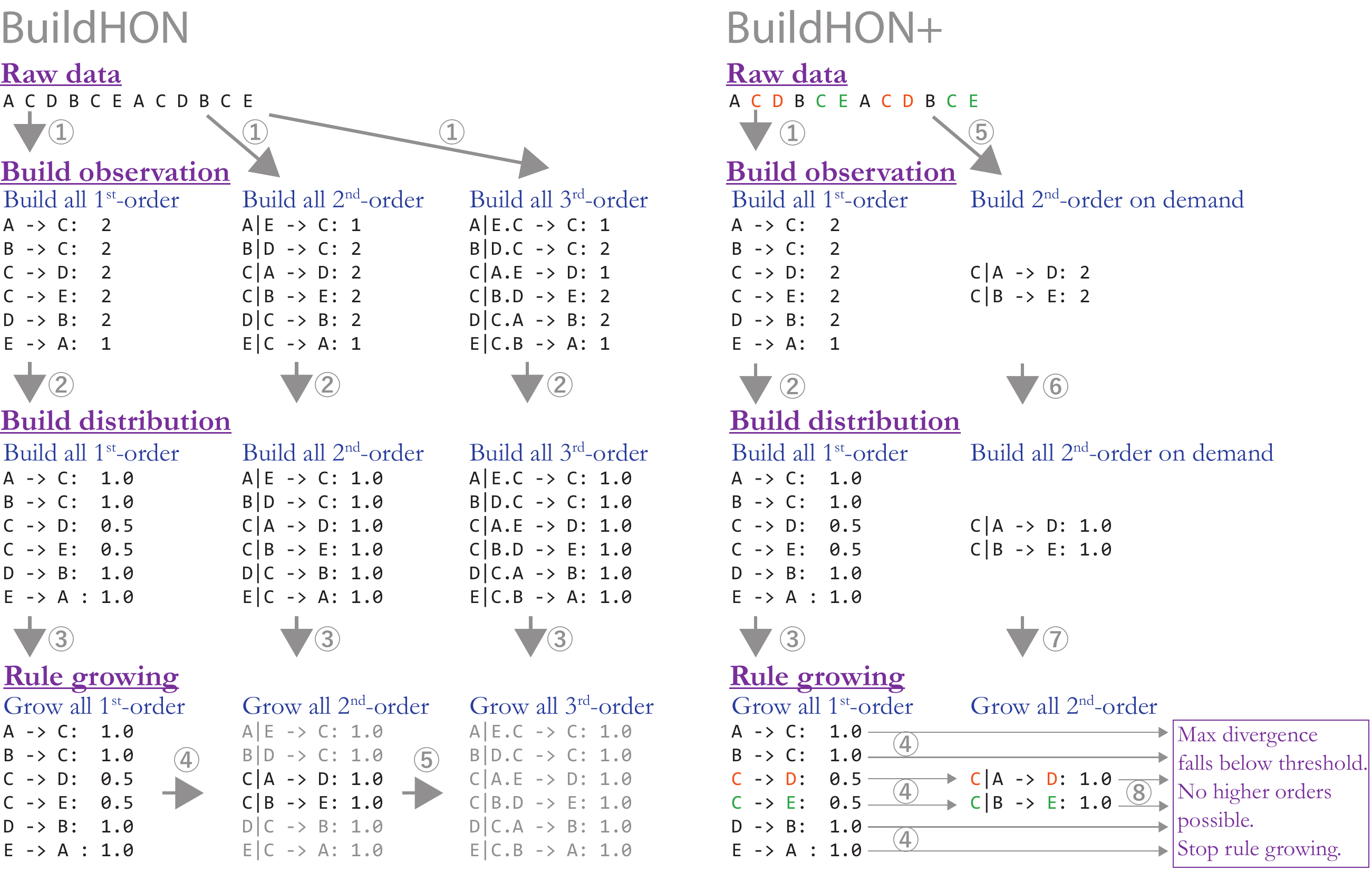}
\caption{\csentence{Comparison of the {\em active} observation construction in \HON (left) and the {\em lazy} observation construction in \HONp (right, with a much smaller search space). Circled numbers represent the order of execution.}}
\label{fig:HON+}
\end{figure*}

The core of \HON is the dependency rule extraction step, which answers whether higher-order dependencies exist in the raw sequential data, and how high the orders are.
The dependency rules extracted are then converted to higher-order nodes and edges as the building blocks of HON.
Rather than deriving a fixed order of dependency for the whole network, the method allows for variable orders of dependencies for more compact representation.
Fig.~\ref{fig:HON+} illustrates the dependency rule extraction step.
\HON first counts the observed n-grams in the raw data (step \raisebox{.5pt}{\textcircled{\raisebox{-.9pt} {1}}}),
then compute probability distributions for the next steps given the current and previous steps (step \raisebox{.5pt}{\textcircled{\raisebox{-.9pt} {2}}}).
Finally test if knowing one more previous step significantly changes the distribution for the next step -- if so, higher-order dependency exists for the path (step \raisebox{.5pt}{\textcircled{\raisebox{-.9pt} {4}}}); this procedure (``rule growing'') is iterated recursively until a pre-defined {\em MaxOrder} (shown here $MaxOrder=3$).
In this example, the probability distribution of the next steps from $C$ changes significantly if the previous step (coming to $C$ from $A$ or $B$) is known (step \raisebox{.5pt}{\textcircled{\raisebox{-.9pt} {4}}}), but knowing more previous steps (coming to $C$ from $E \rightarrow A$ or $D\rightarrow B$) does not make a difference (step \raisebox{.5pt}{\textcircled{\raisebox{-.9pt} {5}}}); therefore, paths $C|A \rightarrow D$ and $C|A \rightarrow E$ demonstrate second-order dependencies.

Formally, the ``rule growing'' process works as follows: for each path (n-gram) $\mathcal{S} = [S_{t-k}, S_{t-(k-1)}, \dots, S_t]$ of order $k$, starting from the first-order $k=1$, assume $k$ is the true order of dependency, which $\mathcal{S}$ has the distribution $D$ for the next step.
Then extend $\mathcal{S}$ to $\mathcal{S}_{ext} = [S_{t-(k+1)}, S_{t-k}, S_{t-(k-1)}, \dots, S_t]$ by adding one more previous step; $\mathcal{S}_{ext}$ has order $k_{ext}=k+1$ and distribution $D_{ext}$.
Next, test if $D_{ext}$ is significantly different than that of $D$ using Kullback-Leibler divergence \cite{kullback1951information} as $\mathcal{D}_{KL}(D_{ext}||D)$, and compare with a dynamic threshold $\delta$ -- if the divergence is larger than $\delta$, order $k+1$ is assumed instead of $k$ for the path $\mathcal{S}_{ext}$.
The dynamic threshold $\delta$ is defined as
$ \delta = \frac{k_{ext}}{\log_2 (1+ Support_{\mathcal{S}_{ext}})}$, so that lower orders are preferred than higher-orders, unless higher-order paths have sufficient support (number of observations).
The whole process is iterated recursively until $MaxOrder$.

\subsubsection{Eliminating all parameters}

The reason for having the {\em MaxOrder} and {\em MinSupport} parameters in \HON is to set a hard stop for the rule growing process, otherwise, it will iterate indefinitely and keep extending $\mathcal{S}$. However, we show that we can pre-determine if extending $\mathcal{S}$ will not produce significantly different distributions, which forms an important basis for \HONp.

\vspace{-0.08in}
\begin{lemma} The significance threshold $ \delta = \frac{k_{ext}}{\log_2 (1+ Support_{\mathcal{S}_{ext}})}$ increases monotonically in rule growing when expanding $\mathcal{S}$ to $\mathcal{S}_{ext}$. 
\end{lemma}

\begin{proof}
  On the numerator, the order $k_{ext}$ of the extended sequence $\mathcal{S}_{ext}$ increases monotonically with the inclusion of more previous steps.
  Meanwhile, every observations of\\ $[S_{t-(k+1)}, S_{t-k},\dots, S_{t-1}, S_t]$ in the raw data can find a corresponding observation of $[S_{t-k}, \dots, S_{t-1}, S_t]$, but not the other way around.
  Therefore, the support of $S_{ext}$, $Support_{S_{ext}}$ $\leq$ $Support_S$ of the lower order $k = k_{ext}- 1$.
 As a result, the denominator decreases monotonically with the rule growing process.
\end{proof}

Given the next step distribution $D=[P_1, P_2, \dots, P_N]$ of sequence $\mathcal{S}$, we can derive an upper-bound of possible divergence:

\begin{equation}
  \begin{split}
  & max(\mathcal{D}_{KL}(D_{ext}||D)) \\&= max(\sum_{i\in D} P_{ext}(i) \times log_2\frac{P_{ext}(i)}{P(i)})\\
   &\leq 1 \times log_2\frac{1}{min(P(i))} + 0 + 0 + \dots\\
   &= -log_2(min(P(i)))
  \end{split}
\end{equation}

The equal sign (maximum possible divergence) is taken iff the least likely option for the next step $P(i)$ in $\mathcal{S}$ becomes the most likely option $P_{ext}(i)=1$ in $\mathcal{S}_{ext}$, and all other options have $P=0$. 
  Therefore, we can test if $-log_2(min(P_{Distr}(i))) < \delta$ holds during the rule growing process; if it holds, then further increasing the order (adding more previous steps) will not produce significantly different distributions, so we can stop the rule growing process and take the last known $k$ (which passed the actual divergence test, not the order which passes the maximum divergence test) as the true order of dependency.
  Note that, the dynamic threshold is chosen heuristically in its current form. This threshold meets our design requirements:  1) enforce higher support for higher-orders, and 2) fast to compute, as it is a frequently used module in the innermost loop.
  
  Furthermore, \HONp no longer requires a {\em MinSupport} parameter. Recall that using {\em MinSupport $>1$} in \HON helps reduce the search space as a crude form of early stopping, with the risk of losing valid higher-order patterns. In \HONp, the dynamic threshold takes care of early stopping without requiring any extra parameter ({\em MinSupport}) to limit the search space.  This parameter is left in the algorithm only for backward compatibility and is set to 1 by default, but does not serve any initial seeding purpose. In other words, MinSupport is not used in \HONp. 

An advantage of this proposed parameter-free approach is that rather than terminating the rule growing process prematurely by the $MaxOrder$ threshold, the algorithm can now extract {\em arbitrarily orders of dependency}.
 
\subsubsection{Scalability for Higher-orders}
\HON builds all observations and distributions up to $MaxOrder$ ahead of the rule growing process (Fig.~\ref{fig:HON+} left).
This procedure becomes prohibitively expensive for big data with high orders of dependencies: to extract sparse tenth order dependencies, \HON needs to enumerate n-grams from first-order to tenth order and compare probability distributions, which already exceeds a personal computer's capacity using a typical real-world data set (see Section~\ref{sec:performance}).

\HONp, on the other hand, uses a {\em lazy construction} of observations and distributions that has a much smaller search space, and can easily scale to arbitrarily high order of dependency.
Specifically, \HONp does not require the counting of the occurrences of n-grams or calculating the distribution of the next steps, until the rule growing step explicitly asks for such information. 

\tb{Example.  }
\HONp first builds all first-order observations and distributions (Fig.~\ref{fig:HON+} right step \raisebox{.5pt}{\textcircled{\raisebox{-.9pt}  {1}}}--\raisebox{.5pt}{\textcircled{\raisebox{-.9pt} {3}}}).
Given that $A\rightarrow C$, $B\rightarrow C$, $D\rightarrow B$, $E\rightarrow A$ all have single deterministic options for the next step with $P=1$, according to $-log_2(min(P_{Distr}(i))) = 0 < \delta$, \HONp knows no higher-order dependencies can possibly exist by extending these bigrams (step \raisebox{.5pt}{\textcircled{\raisebox{-.9pt} {4}}}).
Only the two paths $C\rightarrow D$ and $C\rightarrow E$ will be extended; since the corresponding second-order observations and distributions are not known yet, \HONp {\em selectively} derives the necessary information from the raw data (Fig.~\ref{fig:HON+} right step \raisebox{.5pt}{\textcircled{\raisebox{-.9pt} {5}}}--\raisebox{.5pt}{\textcircled{\raisebox{-.9pt} {7}}}), and finds that the second-order distributions show significant changes.
At this point, both $C|A\rightarrow D$ and $C|B\rightarrow E$ have single deterministic options for the next step, so again, \HONp determines no dependencies beyond second-order can exist (step \raisebox{.5pt}{\textcircled{\raisebox{-.9pt} {8}}}), so the rule growing procedure stops, without the need for further generation and comparison of distributions.

The challenge is how to count the n-gram of interest on demand -- seemingly every on-demand construction requires a traversal of the raw sequential data with the complexity of $\Theta(L)$.
However, given the following knowledge:

\begin{lemma} 
  All observations of the sequence $[S_{t-k-1}, S_{t-k}, \dots,\allowbreak S_{t-1},\allowbreak S_{t}]$ can be found exactly at the current and one preceding locations of all observations of sequence [$S_{t-k}, \dots, S_{t-1}, S_{t}]$ in the raw data.
\end{lemma}

\begin{proof}
Instead of traversing the raw data, we use an {\em indexing cache} to store the locations of known observations, then use that to narrow down higher-order n-gram look-ups.
As illustrated in Fig.~\ref{fig:HON+}, if we cache the locations of $C\rightarrow D$ and $C\rightarrow E$ in the raw sequential data, then $C|A\rightarrow D$ and $C|B\rightarrow E$ can be found at the same locations.

During the rule growing process, if $\mathcal{S}_{ext}$ has not been observed, {\em recursively} check if the lower-order observation is in the indexing cache, and use those cached indexes to perform a fast lookup in the raw data.
New observations from $\mathcal{S}_{ext}$ are then added to the indexing cache.
This procedure guarantees the identification of observations of the previously unseen $\mathcal{S}_{ext}$, and the lookup time for each observation is $\Theta (1)$ when the indexing cache is implemented with hash tables.
\end{proof}


\tb{Complexity analysis.  }
We formally analyze and compare the computational complexity of \HON and \HONp. 

\tb{\HON. }Suppose the size of raw sequential data is $L$, and there are $D_i$ distinct n-grams of order of $i$. All first-order observations (bigrams) take $\Theta(2D_2)$ space, second order observations (trigrams) take $\Theta(3D_3)$ space, and so on; building observations and distributions up to $k^{th}$ order takes $\Theta(2D_2+3D_3+\dots+kD_k)$ storage, with $k$ being the maximum order allowed, because \HON always keeps raising order until $k$ is reached, while keeping all the breadth-first search results for lower orders. with $D_3\geq D_2, D_4  \geq D_3,$ resulting in a complexity of $\Theta(k^2D_2)$. 

If $N$ is the number of unique entities in the raw data, then the time complexity of the algorithm is $\Theta(Nk^2D_2)$. All observations will be traversed at least once, and evaluating if adding a previous step significantly changes the probability distribution of the next step takes up to $\Theta(N)$ time (assuming Kullback-Leibler divergence \cite{kullback1951information} is used).

\tb{\HONp. } Assume there are $R_i$ distinct n-grams that are {\em exactly} of order $i$. By definition, we have $R_i \leq D_i$. Therefore, \HONp's space complexity is $\Theta(2R_2+3R_3+\dots+tR_t)$ (including observations, distributions, and the indexing cache) where $R_k$ is the {\em exact} number of higher-order dependency rules for order $k$.
Note that, $R_k\leq L$, but it is not necessarily $R_{(i+1)}\geq R_{(i)}$. Also $t\leq k$. 


In practice, what makes \HONp different from \HON is its sensitivity to the underlying data. If the dataset contains very few non-significant n-grams up to maximum specified order, the space complexity of \HONp would not be very different from \HON. However, for very noisy data $(D_i >> R_i)$ or data with an actual order much smaller than the specified maximum order $(t << k)$, the space complexity of \HONp would be significantly smaller than \HON. The same applies to time complexity: while \HON has $\Theta(Nk^2D_2)$, \HONp has $\Theta(N(2R_1+3R_2+\dots))$.
A side-by-side comparison between \HON and \HONp in running time and memory consumption on a real-world data set is provided in Section~\ref{sec:performance}.


\subsection{Higher-order Anomaly Detection}

\tb{Definition.  }The procedure of a network-based anomaly detection method takes the sequential data, $\mathcal{S} = [S_1, S_2, \dots,\allowbreak S_T]$ which is divided into $T$ time windows $t \in [1, T]$ as the input.
In each time window, the sequential data is represented as a network, i.e., $S_i \rightarrow G_i$, yielding a dynamic network $\mathcal{G} = [G_1, G_2, \dots, G_T]$ composed of the sequence of networks. The dynamic network $\mathcal{G}$ is then used to find the change point(s) $t \in [1, T]$ when $G_{t}$ is significantly different from $G_{t-1}$.
The difference between networks in neighboring time intervals, i.e., $d_t=\mathcal{D}(G_{t-1}, G_{t})$, can be quantified by network distance metrics $\mathcal{D}$ (e.g., \cite{shoubridge2002detection,kraetzl2006modality,pincombe2005anomaly}).
Then the problem of anomaly detection in networks reduces to anomaly detection in the time series of $[d_2, d_3, \dots, d_{T}]$.
Next, to determine if the network difference $d_t$ is significantly high, straightforwardly, if $d_t$ is larger than a fixed threshold $\Delta$, $G_{t}$ is anomalously different than $G_{t-1}$. 
Another more robust way is to establish the norm of network differences by computing the running average and standard deviation of network differences in the last $k$ time intervals, the null hypothesis being $d_t$ not significantly large; if $d_t$ deviates from the running average by two standard deviations, the null hypothesis is rejected and time $t$ is considered a change point.

Existing network-based anomaly detection methods mostly differ at the network distance calculation step. However, for the $S_i \rightarrow G_i$ step, i.e., where raw sequential data is represented as networks, existing methods all use FON as $G$ to represent the underlying sequential data $S$, by counting the occurrences of pairs (bigrams) as edge weights in the network.
Here, we propose to use the higher-order network (HON) that selectively embeds n-grams for the $S_i \rightarrow G_i$ step.
HON, using \HONp, keeps all structures of FON, and when higher-order dependencies exist in the raw sequential data, it splits a node into multiple nodes representing previous steps. 
We show that certain types of anomalies will remain undetected for all existing network-based anomaly detection methods using FON, but can be revealed by using HON.

\begin{figure*} 
\centering
\includegraphics[width=.78\linewidth]{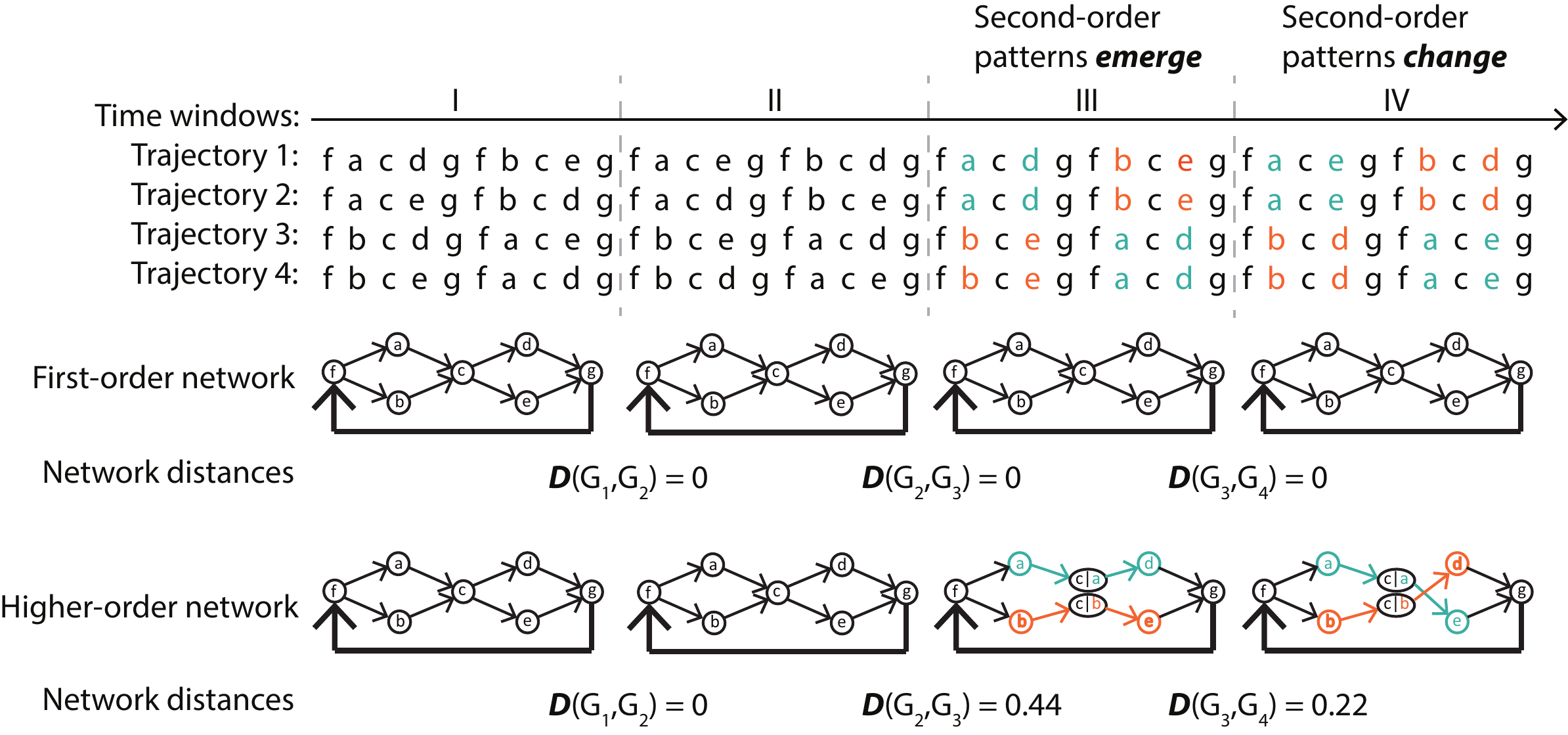}
\caption{\csentence{Comparing anomaly detection on taxi trajectories based on the first-order dynamic network and the higher-order dynamic network.}}
\label{fig:AnomalyIdea}
\end{figure*}

\tb{Example. } 

Fig.~\ref{fig:AnomalyIdea} illustrates a side-by-side comparison of FON and HON in the network representation step.
Suppose there are four taxi trajectories in the raw data. In time window I, taxis in location $c$ randomly navigate to $d$ or $e$, regardless if the taxis came to location $c$ from $a$ or $b$.
In this time window, HON is identical to FON and there are no higher-order dependencies.
In time window II, the traffic patterns are randomly shuffled, and the pairwise traffic between pages $a, b, c, d, e$ remains the same as time window I. Neither FON nor HON shows changes.

In time window III, second-order patterns {\em emerge}: all taxis that had navigated from $a$ to $c$ go to $d$, and all taxis from $b$ to $c$ go to $e$. Since the aggregated traffic from $c$ to $d$ and $e$ remains the same, the FON remains {\em exactly the same}, missing this newly emerged pattern.
In contrast, HON uses additional higher-order nodes and edges to capture higher-order dependencies: node $c$ is now splitted into a new node $c|a$ (representing $c$ given the last step being $a$) and node $c|b$ (representing $c$ given the last step being $b$). The path $a\rightarrow c \rightarrow d$ now becomes $a \rightarrow c|a \rightarrow d$; the edge $c \rightarrow e$ rewired similarly.
Therefore, the emergence of the second-order pattern in the raw data is reflected by the non-trivial changes in the topology of HON.
If we use the weight distance \cite{shoubridge2002detection} defined as
\begin{equation}
\label{eq:weight}
\mathcal{D}(G, H) = |E_G\cup E_H|^{-1}\sum\limits_{u,v\in V}\frac{|w^G(u,v)-w^H(u,v)|}{max\{w^G(u,v),w^H(u,v)\}}
\end{equation}
with $w$ being the edge weights and $|E|$ being the total number of edges,
due to the complete changes in four out of the nine edges on HON, the network distance $\mathcal{D}(G_2, G_3) = 0.44 > 0$, successfully captures this {\em higher-order anomaly} (a significant change in higher-order navigation patterns).

In time window IV, the second-order navigation pattern {\em changes}: all taxis that navigated from location $a$ to $c$ now visit $e$ instead of $d$, and all from $b$ to $c$ now visit $d$ instead of $e$.
Since the pairwise traffic from $c$ to $d$ and $e$ remains the same, FON remains the same.
However, HON captures the changes with two edge rewirings: now $c|a \rightarrow e$ and $c|b \rightarrow d$, resulting in $\mathcal{D}(G_3, G_4) = 0.22 > 0$.

In brief, FON is an oversimplification of sequential data produced by complex systems, and conventional network-based anomaly detection methods that use FON may fail to capture the emergence and changes of higher-order navigation patterns. If HON is used instead, without changes to distance metrics, existing methods can capture these previously ignored anomalies.

\subsubsection{Distance Metrics}

After successful construction of HON (using \HONp) we apply  five network distance measures to detect anomalies. 
\begin{compactenum}
\item \tb{Weight distance. } This metric was introduced earlier (Equation~\ref{eq:weight}).

\item \tb{Maximum common subgraph (MCS). } The MCS distance is defined similarly to the weight distance in Equation~\ref{eq:weight} but operates on MCS ~\cite{shoubridge2002detection}:
\begin{equation}
\mathcal{D}(G, H) = |E_G\cap E_H|^{-1}\sum\limits_{u,v\in V}\frac{|w^G(u,v)-w^H(u,v)|}{max\{w^G(u,v),w^H(u,v)\}}
\end{equation}
\item \tb{Modality. } This distance function can be defined as follows~\cite{kraetzl2006modality}:
\begin{equation}
\mathcal{D}(G, H)= ||\pi(G)-\pi(H)||
\end{equation}
where $\pi(G)$ and $\pi(H)$ are the Perron vectors of graphs $G$ and $H$, respectively. 
\item \tb{Entropy graph distance. } This can be defined as follows~\cite{pincombe2005anomaly}:
\begin{equation}
\mathcal{D}(G, H)= E(G)-E(H)
\end{equation}
where $E(*)$ is the entropy measure of the edges:
\begin{equation}
E(*)=\sum\limits_{e\in E_*}\widetilde{W}_*^e-\sum\limits_{e\in E_*}ln~\widetilde{W}_*^e
\end{equation}

and:
\begin{equation}
\widetilde{W}_*^e=\frac{W_*^e}{\sum\limits_{e\in E_*}W_*^e}
\end{equation}
is the normalized weight for edge e.
\item Finally, we also use the \tb{spectral distance}, which is defined as~\cite{pincombe2005anomaly}:

\begin{equation}
    \mathcal{D}(G, H)=\sqrt{
\frac{\sum\limits_{i=1}\limits^k(\lambda_i-\mu_i)^2}{min(\sum\limits_{i=1}\limits^k\lambda_i^2,\sum\limits_{i=1}\limits^k\mu_i^2)}}
\end{equation}

where $\lambda_i$ and $\mu_i$ represent the eigenvalues of the Laplacian matrix for graph $G$ and $G$, respectively.
\end{compactenum}
Note that, in order to calculate network distance in HON, all higher-order nodes are treated as first-order ones. That is, a change from $D|B,C \rightarrow E$ to $D|B,C,A \rightarrow E$ results in total removal of $D|B,C$ and new addition of node $D|B,C,A$. The reason is that in many cases, anomalous patterns result in a change of higher-order patterns. It is desirable that the anomaly detection method detects the ``emergence'', ``change'' and ``dissipation'' of higher-order patterns. We leave the task of classifying different higher-order anomalies for future work.

\section{Results}

In this section, we first compare \HONp with \HON in terms of running time and memory consumption on real-world data of various sizes and multiple orders of dependency. Next, we present the anomaly detection results.

For the anomaly detection experiments, we first construct a large-scale synthetic taxi movement data with 11 billion movements and variable orders of dependencies, and show that five existing anomaly detection methods based on FON collectively fail to capture anomalous navigation behaviors beyond first-order, while using our framework, all methods show significant improvements.

We also demonstrate HON on real-world taxi trajectory data, showing its ability in capturing the higher-order anomaly signals and revealing the exact location of anomalies.

\label{sec:performance}
\begin{figure}[!htp]
\centering
\includegraphics[width=.9\linewidth]{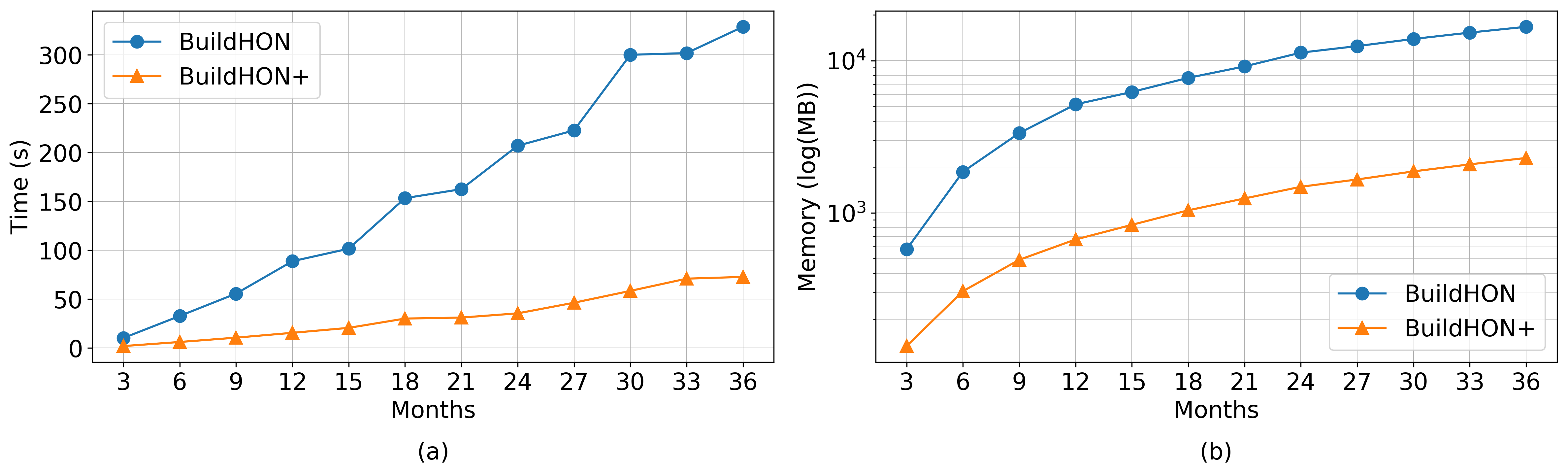}
\caption{\csentence{\HON is highly sensitive to the size of the data.  For the maximum data size, \HON takes 4.5 times longer than \HONp to run (a), and requires approximately 7.2 times more memory than \HONp (b) . We set  {\em MaxOrder=15} for BuildHON.}}
\label{fig:benchmark_1}
\end{figure}

\subsection{Scalability Analysis: Performance improvement of \HONp over \HON}
To highlight the scalability advantage of \HONp, instead of the taxi data or the synthetic data (which demonstrates up to third order of dependency), we use the same shipping trajectories data as used in the HON paper ~\cite{xu2016representing}. This data was shown to demonstrate dependencies of more than the fifth-order due to ships' cyclic movement patterns. It consists of up to three years of shipping data (between May ${1^{st}}$, 1997 and April ${30^{th}}$, 2003), aggregated over 3-months intervals. The smallest and largest data contains 372,500 and 4,721,936 voyages, respectively.

For a fair comparison, we use the Python implementation for both \HONp and \HON. Both implementations run single-threaded on the same Linux machine (Intel Quad 16-core @ 2.10GHz, 128 GB RAM). \HONp is parameter-free (no limit to the maximum order, optional {\em MinSupport} = 1) and does not require further configuration. We set {\em MinSupport} = 1 and {\em MaxOrder} = 15 for \HON. We start with the first 3-months of the data and aggregate the trajectories over the next 6 months, 9 months, and so on. Fig.~\ref{fig:benchmark_1} illustrate the time and memory usage of both algorithms as the size of the data increases. We observe that \HON is highly sensitive to the size of the data. For the maximum data size, \HON requires approximately 7.2 times more memory than \HONp and takes 4.5 times longer to run.

\begin{figure}[!htp]
\centering
\includegraphics[width=.5\linewidth]{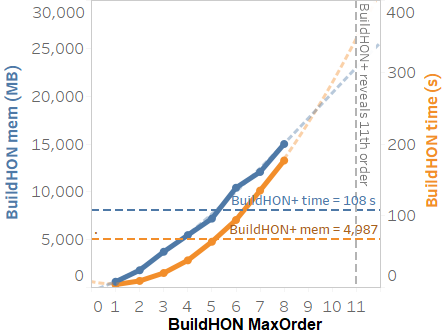}
\caption{\csentence{Given the same data ~\cite{xu2016representing}, \HONp extracts up to $11^{th}$ order in 1/3 run-time and 1/5 memory of \HON. We set {\em MaxOrder=11} for \HON.}}
\label{fig:benchmark_2}
\end{figure}

We further analyze the run time and memory usage of both algorithms on the same shipping dataset to analyze the effect of setting different values for {\em MaxOrder}. For this experiment, we use one year of data which consists of 3,415,577 voyages between May ${1^{st}}$, 2012 and April ${30^{th}}$, 2013.

We set {\em MinSupport} = 1 for \HON, and gradually increase {\em MaxOrder} from the first-order. Same as above,  \HONp does not require further configuration.
\HONp was able to find up to $11^{th}$ order within 2 minutes, with a peak memory usage less than 5GB, as the reference lines displayed in Fig.~\ref{fig:benchmark_2}. In comparison, \HON already exceeds the running time and memory consumption of \HONp at $6^{th}$ order, reaches the physical memory limit at $8^{th}$ order, and would need about 22 GB memory and 6 minutes (3x time and 5x memory than \HONp) to achieve the same results as \HONp can. Both implementations run single-threaded on the same laptop (Intel i7-6600U @ 2.60GHz, 16GB RAM, SSD).

\subsection{Anomaly Detection: Large-scale Synthetic Taxi Movements}

We first use the synthetic data with known higher-order anomalies to test the effectiveness of the HON-based anomaly detection framework. 
With synthetic data, we know exactly when, where, and what types of anomalies exist.
To begin with, we assume 100,000 taxis are navigating through a 10x10 grid with cells numbered from 00 to 99. At each timestamp, every taxi moves 100 steps, resulting in 10,000,000 movements. 

Our goal is to synthesize input sequences with variable orders of taxi navigating patterns.
We start from the basic case where all taxis navigate randomly, then gradually add or change first-order and higher-order navigation rules, and see if the proposed method can successfully identify these anomalies.

\begin{figure*}
\centering
\includegraphics[width=\linewidth]{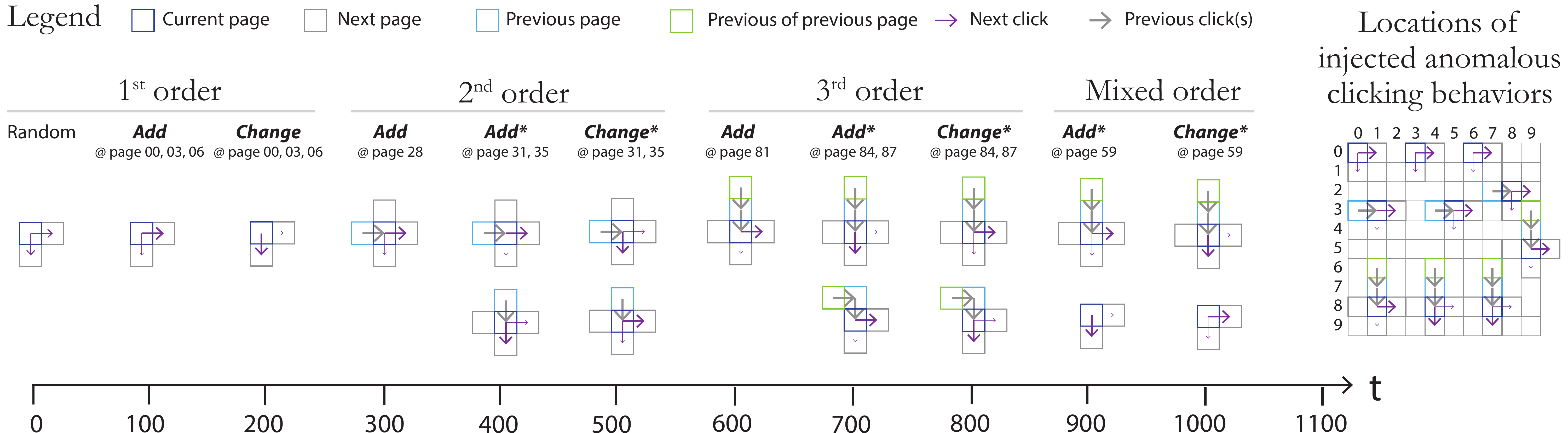}
\caption{\csentence{Synthetic taxi movement data: variable orders of navigation patterns on 100 cells as a 10x10 grid.}}
\label{fig:AnomalySynthetic}
\end{figure*}

For each of the following 11 cases, we maintain the taxi navigation behavior for 100 time windows.
In total, we generate 11,000,000,000 taxi movements for the subsequent anomaly detection task.
The full process to synthesize the input trajectories is illustrated in Fig.~\ref{fig:AnomalySynthetic}.

\tb{Initial random movement case.  } At $t=[0,99]$, each taxi has a 50\% chance of navigating to the cell on the right and 50\% chance of navigating down in each move.\\
\tb{Emergence of the first-order dependency.  }
At $t=[100,199]$, we impose the following first-order rule of movement: all taxis coming to cell 00, 03 and 06 will have a 90\% chance of moving to the right and 10\% chance of moving down in the next step.
This new rule incurs a significant change of first-order traffic at $t=100$ between pairs of cells 00--01, 00--10, 03--04, 03--13, 06--07 and 06--16.
The locations of these dependency rules are highlighted on the right of Fig.~\ref{fig:AnomalySynthetic}.

\begin{figure*}
\centering
\vspace{-.05in}
\includegraphics[width=0.5\linewidth]{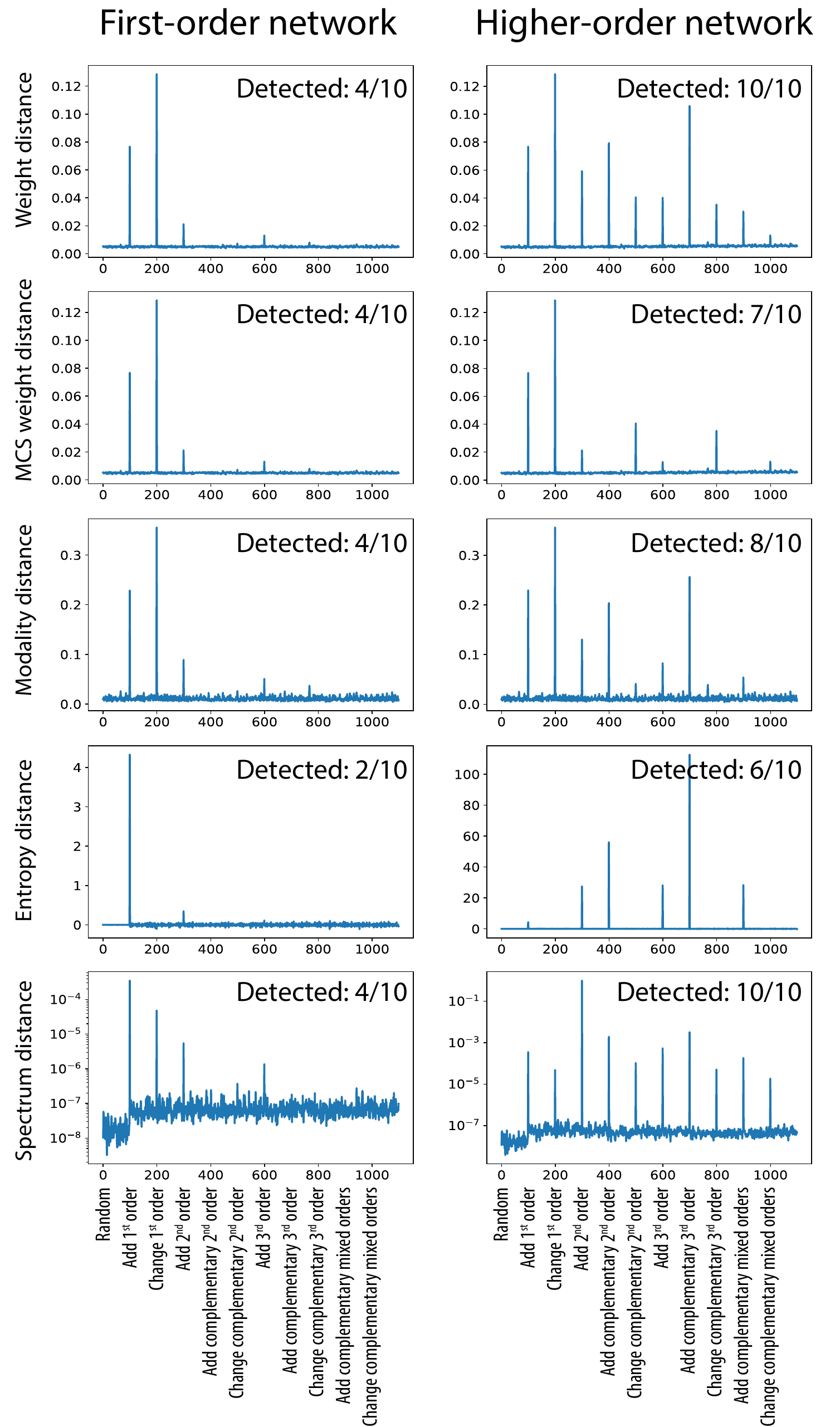}
\caption{\csentence{Methods that use FON collectively fail to capture anomalous navigation behaviors beyond first-order no matter what distance metric is used, but all show signals when HON is used instead.}}
\label{fig:AnomalySyntheticResults}
\end{figure*}

\tb{Change of the first-order dependency.  }
At t = [200,299], we change the existing first-order movement rules: all taxis coming to cell 00, 03 and 06 will now have a 90\% chance of moving down in the next step, and a 10\% chance of moving right. This change at t = 200 should also be reflected in both FON and HON.

\tb{Emergence of second-order dependency.  }
At $t=[300,399]$, we keep the previous first-order rules and impose a new second-order rule: all taxis coming from cell 27 to 28 will have a 90\% chance of moving to the right in the next step, and a 10\% chance of moving down.
This change at $t=300$ not only introduces new higher-order dependencies, but also slightly influences first-order traffic (traffic of $27\rightarrow 28 \rightarrow 29/38$ changes from 1:1 to 7:3).

\tb{Emergence of complementary second-order dependencies.  }
At $t=[400,499]$, we impose a pair of new second-order rules: (1) all taxis coming from cell 30 to 31 (and 34 to 35) will have a 90\% chance of moving to the right in the next step, and a 10\% chance of moving down; (2) all taxis coming from page 21 to 31 (and 25 to 35) will have a 90\% chance of moving down, and a 10\% chance of moving right.
The combined effect of these two new complementary second-order dependencies at $t=400$ is that the first-order taxi traffic from cell 31 and 35 remains unchanged.

\tb{Change of complementary second-order dependencies.  }
At $t=[500,599]$, we flip the rules for the complementary second-order dependencies: (1) all taxis coming from cell 30 to 31 (and 34 to 35) will have a 90\% chance of moving down, and a 10\% chance of moving right; (2) all taxis coming from page 21 to 31 (and 25 to 35) will have a 90\% chance of moving right, and a 10\% chance of moving down.
At $t=500$ the first-order taxi traffic from cell 31 and 35 still remains unchanged.

\tb{Emergence of third-order dependency.  }
At $t=[600,699]$, we impose a new third-order rule: all taxis coming from cell 61 through 71 to 81 will have a 90\% chance of moving to the right in the next step, and a 10\% chance of moving down.
This introduction of third-order dependencies at $t=600$ also slightly influences the first-order traffic (from 1:1 to 3:2).

\tb{Emergence of complementary third-order dependencies.  }
At $t=[700,799]$, we impose a pair of new third-order rules: (1) all taxis coming from cell 64 through 74 to 84 (and 67 through 77 to 87) will have a 90\% chance of moving to the right in the next step, and a 10\% chance of moving down; (2) all taxis coming from 73 through 74 to 84 (and 76 through 77 to 87) will have a 90\% chance of moving down, and a 10\% chance of moving right.
Here at $t=700$ first-order traffic does not change when these two complementary third-order dependencies are introduced together.

\tb{Change of complementary third-order dependencies.  }
At $t=[800,899]$, we flip the rules for the complementary third-order dependencies.
First-order traffic at $t=800$ again remains unchanged.

\tb{Emergence of complementary mixed-order dependency.  }
At $t=[900,999]$, we impose a new third-order rule and a first-order rule: (1) all taxis coming from cell 39 through 49 to 59 will have a 90\% chance of moving to the right in the next step, and a 10\% chance of moving down; (2) all taxis at cell 59 will have 11/30 chance of moving right and 19/30 chance of moving down.
At $t=900$ first-order traffic does not change, because the influence of the new third-order rule on pairwise traffic is canceled by the new first-order rule.

\tb{Change of complementary mixed-order dependency.  }
At $t=[1000,1099]$, we flip the rules for the mixed-order dependencies.
First-order traffic at $t=1000$ remains unchanged.

\subsubsection{Results}
For all five distance metrics, we present a side-by-side comparison between anomaly detection results using FON and HON in Fig.~\ref{fig:AnomalySyntheticResults}.
The Y-axis shows the graph distances between neighboring time windows; given that we have injected 10 anomalous movement patterns at $t=[100, 200, \dots, 1000]$, we should expect to see 10 ``spikes'' in graph distances.

Methods using FON can detect at most 4 out of the 10 anomalies: the addition and changes in first-order movement patterns ($t=100, t=200$), the addition of second-order ($t=300$), and the addition of third-order ($t=600$) movement patterns. 
Because the latter two cases also slightly change the first-order traffic, FON does reflects the changes, but the spikes incurred are not as significant as when changes are made directly to first-order rules.
For the other six cases, all five distance metrics appear as if there are no anomalies, as long as they rely on FON topology.

In contrast, methods using HON (1) capture all first-order anomalies ($t=100, t=200$); (2) show stronger signals for the addition of second-order and third-order rules ($t=300, t=600$) because not only the first-order traffic is changed but \HONp also creates additional higher-order nodes and edges for higher-order dependencies; (3) capture the six additional cases where higher-order movement patterns are changed but first-order traffic remains the same.
Here the topological changes of HON are best reflected with weight distance and spectrum distance (detecting 10/10 anomalies); MCS weight method misses the addition of higher-order nodes and edges ($t=400, 700, 900$) because those topological changes are excluded from common subgraphs; entropy method misses changes in edge weights ($t=200, 500, 800, 1000$), also because by definition a swap in edge weights do not change a graph's entropy.
Nevertheless, all these distance metrics are able to identify more types of anomalous signals simply by using HON instead of FON, with no changes to these distance metrics.
In other words, \HONp can be plugged into existing network-based anomaly detection methods directly, and extend their ability in detecting higher-order anomalies.

\begin{figure*}
\centering
\includegraphics[width=.8\linewidth]{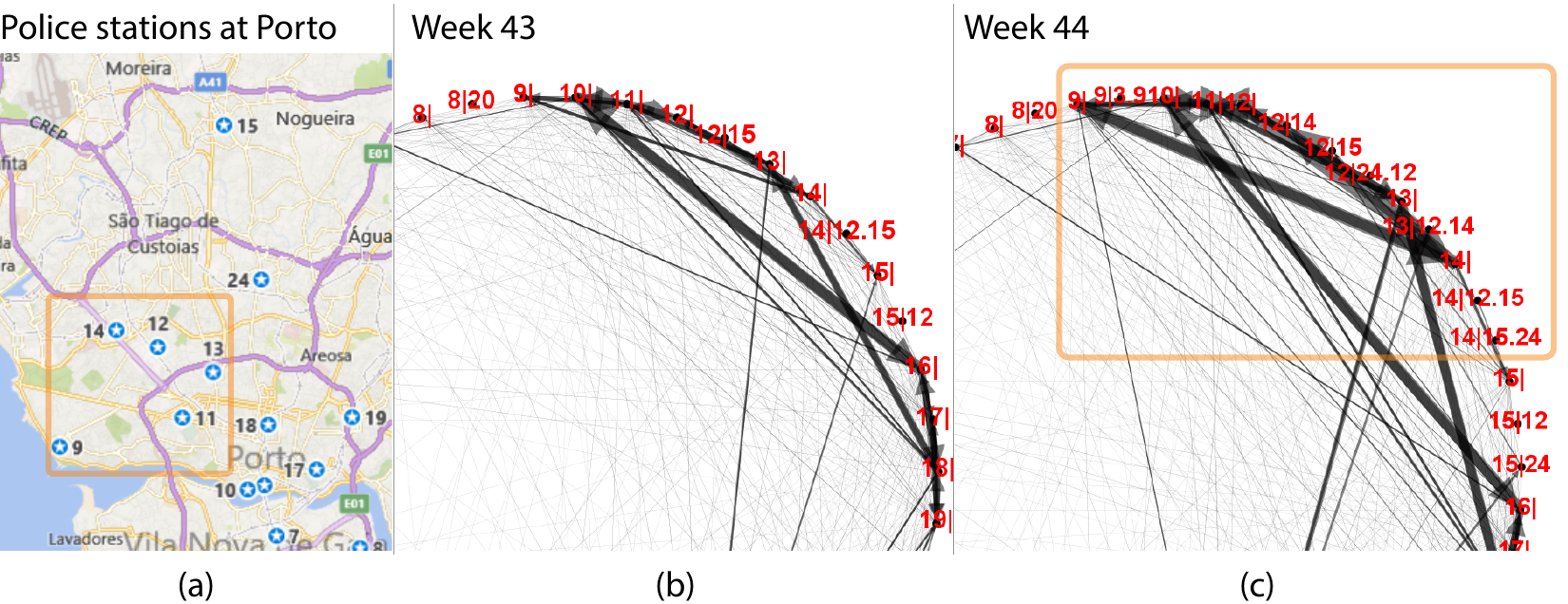}
\caption{\csentence{ (a) Labeling of police stations in urban areas of Porto. (b) and (c) the emergence of higher-order traffic patterns in week 43 and 44 (“Burning of the Ribbons” festival) captured by HON, corresponding to the highlighted region in (a).}}
\label{fig:PortoCase}
\end{figure*}

\begin{figure*}
\centering
\includegraphics[width=.6\linewidth]{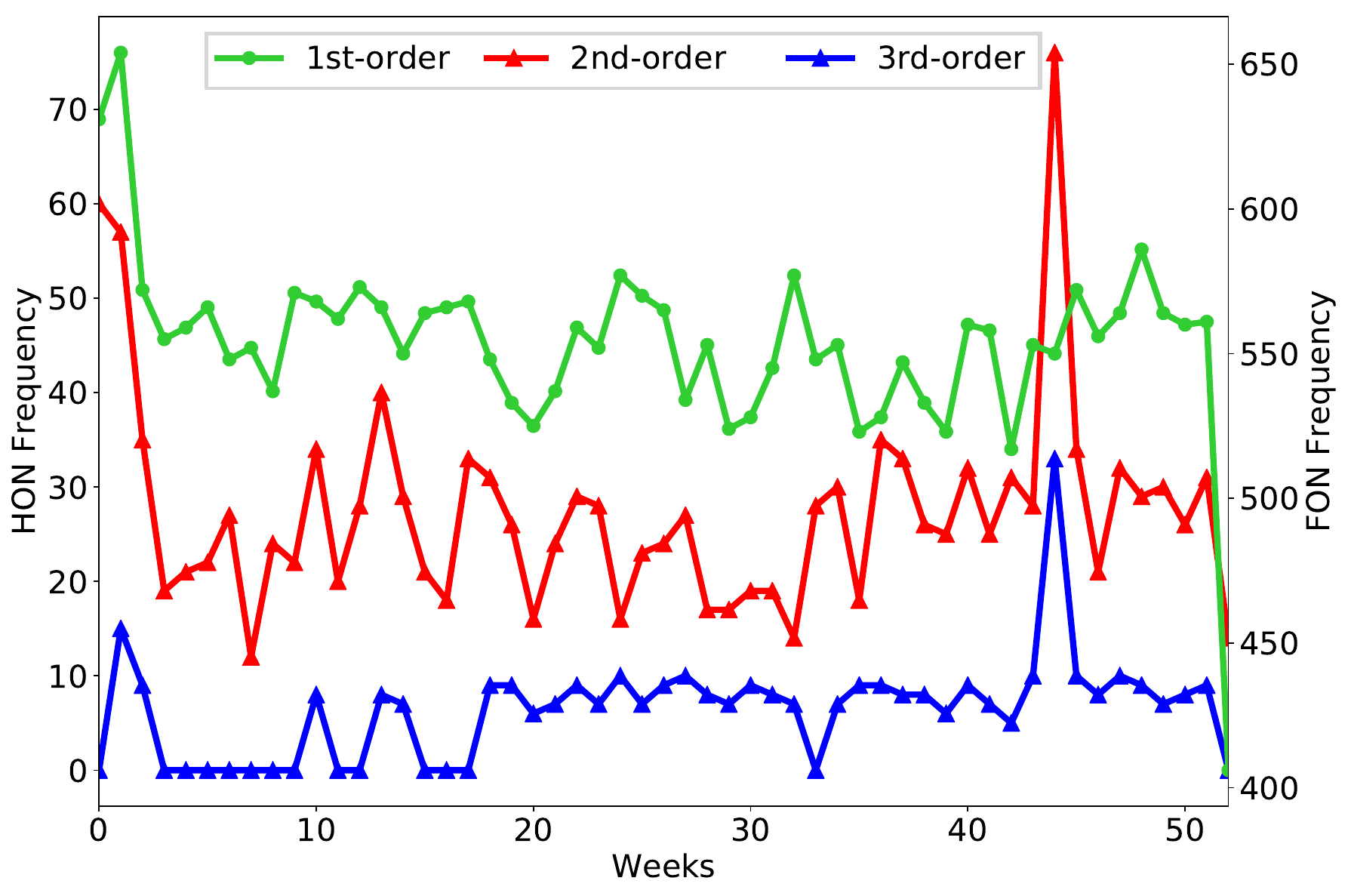}
\caption{\csentence{ Variation of number of first-order, second-order and third-order nodes in HON of the taxi data of Porto. The anomalous traffic patterns result in a significant change in the number of second and third-order nodes, but not the first-order nodes.}}
\label{fig:OrderFreq}
\end{figure*}

\subsection{Anomaly Detection: Real World Taxi Data}
We use the ECML/PKDD 2015 challenge data\footnote{\url{http://www.geolink.pt/eomlpkdd2015-challenge/data set.html}}, which contains one year (Jul. 1, 2013 to Jun. 30, 2014) of all the 442 taxi GPS trajectories in Porto, Portugal.
The coordinates of each taxi were collected every 15 seconds. 
To discretize the geolocation data into points of interest that are representative of population density, we map all coordinates to the nearest 41 police stations (Fig.~\ref{fig:PortoCase}). As a pre-processing step and to void introduction of bias / noise, we removed the taxis that have been idle for more than 5 days because that can arise due to data collection errors (on average 5.29\% of the trajectories were removed). The highlighted box in Fig.~\ref{fig:PortoCase} indicates the detection of anomalies.  Fig.~\ref{fig:OrderFreq} shows the week to week difference in higher-order dependencies, yielding in 52 time windows and 442 trajectories of points of interest. We consider both FON and \HONp with a fixed maximum order of 2 and \HONp with a variable higher-order (discovered to be 3 by the algorithm). We show that \HONp when allowed to discover the maximum order, results in the highest indication of potential anomalies. 

Note that, the choice of time-window is quite data-dependent. We initially attempted daily time-windows but noticed that the weekly fluctuation patterns (weekday commute traffic vs weekend recreational traffic) dominate any other signals. Besides, daily time windows have sparser observations, resulting in a very sparse network for each time step. On the other hand, because anomalous traffic patterns usually last for no more than a few days, using monthly aggregation will dilute the signal and result in too coarse a granularity. 

\begin{figure*}
\centering
\includegraphics[width=0.95\linewidth]{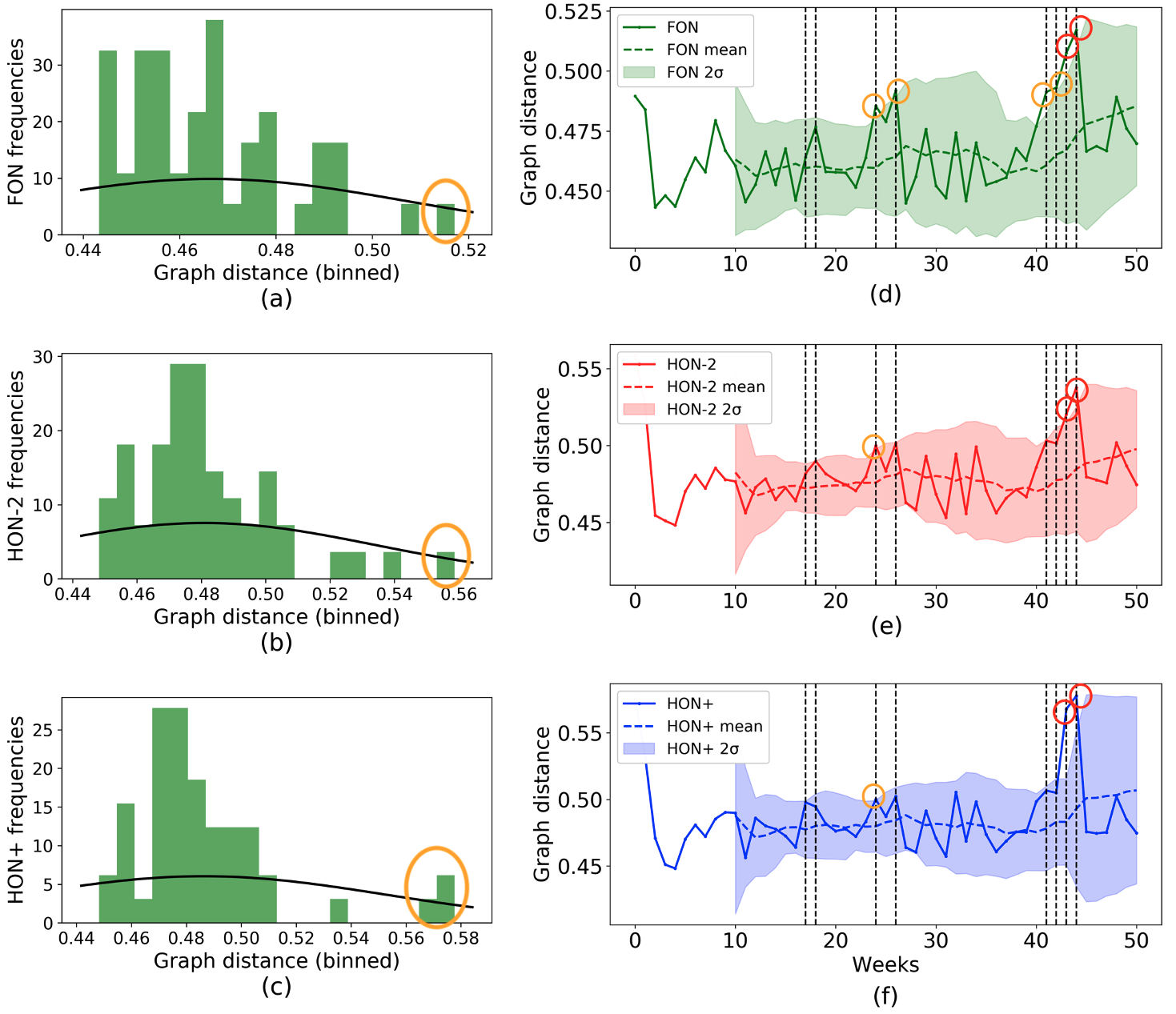}
\caption{\csentence{The anomalous traffic patterns are more noticeable in HON+ (\tb{c}), while they are not as significant in HON-2 (\tb{b}) and FON (\tb{a}). (\tb{d}) Anomaly detection results on the dynamic network of FON and HON-2. (\tb{e}) Anomaly detection results on the dynamic network of HON-2 and HON+.}}

\label{fig:AnomalyPorto}
\end{figure*}

\subsubsection{Graph distance analysis}
We compare the 52 networks for FON, {\em MaxOrder} of 2 as a constraint for \HONp (indicated as HON-2), and no {\em MaxOrder} constraint on \HONp (indicated as HON+) in Fig.~\ref{fig:AnomalyPorto}. Our goal is to see the improvements afforded by allowing \HONp to automatically discover the requisite higher-order for a given data, versus specifying the maximum order of 2 using \HON and the FON representation.  We compute the graph distances (using weight distance) for neighboring time windows. The histograms of graph distances for each network is shown in  Fig.~\ref{fig:AnomalyPorto} (a), (b), and (c).
We also compute the running average and standard deviation using the graph distances in the previous 10 weeks, with the null hypothesis as ``the network is not significantly different if the graph distance does not deviate more than $2\sigma$ from the mean''.
Fig.~\ref{fig:AnomalyPorto} (e) shows the comparison of HON and FON and Fig.~\ref{fig:AnomalyPorto} (d) shows the comparison of HON+ and HON-2. While the trend of HON+ resembles that of HON-2 and FON, the graph distances in weeks 43 and 44 are particularly more significant in HON+ than HON-2 and FON (HON-2 offers more significance over FON as well).
Such differences are also indicated in the histograms of graph distances in Fig.~\ref{fig:AnomalyPorto}(a),(b), and (c), where the orange circles highlight the same anomalous signals, which is observable in HON+, while it is not as significant in FON and even HON-2.

We focus on the case of week 43 and 44 to understand why HON+ produces a stronger signal than HON-2 and FON in this time window.
We notice that Porto's second most important festival, ``Burning of the Ribbons'', lasts from May 2 to May 9 in 2014 and falls within the end of week 43 and the entire week 44 of our study.
The festival involves parades, road closures, and is popular among tourists, which could be the underlying reason for the changes in taxis' movement patterns.
After plotting the traffic HON+ of week 43 and week 44 in Fig.~\ref{fig:PortoCase}(b) and (c), we notice that multiple higher-order nodes and edges emerge in these weeks, indicating the emergence of higher-order traffic patterns.
The newly emerged higher-order patterns correspond to police stations labeled from 9 to 14, which is where the event's main venue (Queimódromo in the City Park) and participating universities are located.\footnote{\url{http://www.maiahoje.pt/noticias/ler-noticia.php?noticia=577}}
We further compared the fluctuations in the number of higher-order nodes (obtained from HON+) in Fig.~\ref{fig:OrderFreq}. We notice that the number of first-order nodes does not change significantly, while the number of second and third-order nodes shows a sharp change in week 44 of the data. FON (although showing deviation from average at week 44) does not capture the change in additional higher-order nodes, and HON-2 does not capture the change in third-order nodes. HON+ is more effective in deciphering the anomalous signal. This analysis shows the importance of including variable and higher-order dependencies for anomaly detection, and the applicability of \HONp in discovering the appropriate orders given the data. Depending on the data, the {\em MaxOrder} value required for accurate detection of anomalies can be different. \HONp removes this dependency, ensuring accurate detection of changes in the network. 

\begin{table*}
\small 
\centering
\begin{tabular}{|l|lr|lr|lr|lr|}
\hline
    & \multicolumn{2}{c|}{\textbf{FON before noise}} & \multicolumn{2}{c|}{\textbf{FON after noise}} & \multicolumn{2}{c|}{\textbf{HON+ before noise}}  & \multicolumn{2}{c|}{\textbf{HON+ after noise}}\\
    \hline
                        &Week	        &Value	        &Week	        &Value	&Week	        &Value	&Week	            &Value\\
                        \hline
	                    &24	            &0.008	        &17	            &0.006	&24	            &0.0001	&41	                &0.003  \\
                    	&26	            &0.003	        &18	            &0.008	&\textbf{43}	&\textbf{0.052}	            &\textbf{43}    &\textbf{0.044}  \\
                    	&41	            &0.009	        &26	            &0.002	&\textbf{44}	&\textbf{0.026}	            &\textbf{44}	&\textbf{0.025}  \\
                    	&42	            &0.002	        &41	            &0.020	&--	            &--	    &--	                &--     \\
                    	&\textbf{43}	&\textbf{0.010}	&42	            &0.004	&--	            &--	    &--	                &--     \\
                    	&\textbf{44}	&\textbf{0.008}	&\textbf{44}	&\textbf{0.003}	        &--	    &--	             &--	    &--     \\
\hline
TP;FP	    &2;4	&	    &1;5	    &	    &2;1	    &	    &2;1      &       \\	
\hline
Precision   &0.333	&	    &0.167      &	    &0.667      &		&0.667    &       \\
\hline
Recall      & 1     &       &0.5        &       &1          &       &1          &\\
\hline
\end{tabular}
\caption{\textbf{Represents all the data points in which the graph distance (using FON and HON+) falls outside the $2\sigma$  threshold. The column ``value'' indicates the difference between the graph distance (obtained from FON or HON+) and the closest $2\sigma$  threshold (with respect to the moving average). TP refers to true positive values (correct anomalies), which are marked as bold. FP refers to false positive values. HON+ correctly detects anomalies, results in much lower false positives, and is more robust in presence of noise. }}
\label{tab:values}
\end{table*}

\subsection{Robustness to noise}
\begin{figure}
    \centering
    \includegraphics[width=0.98\linewidth]{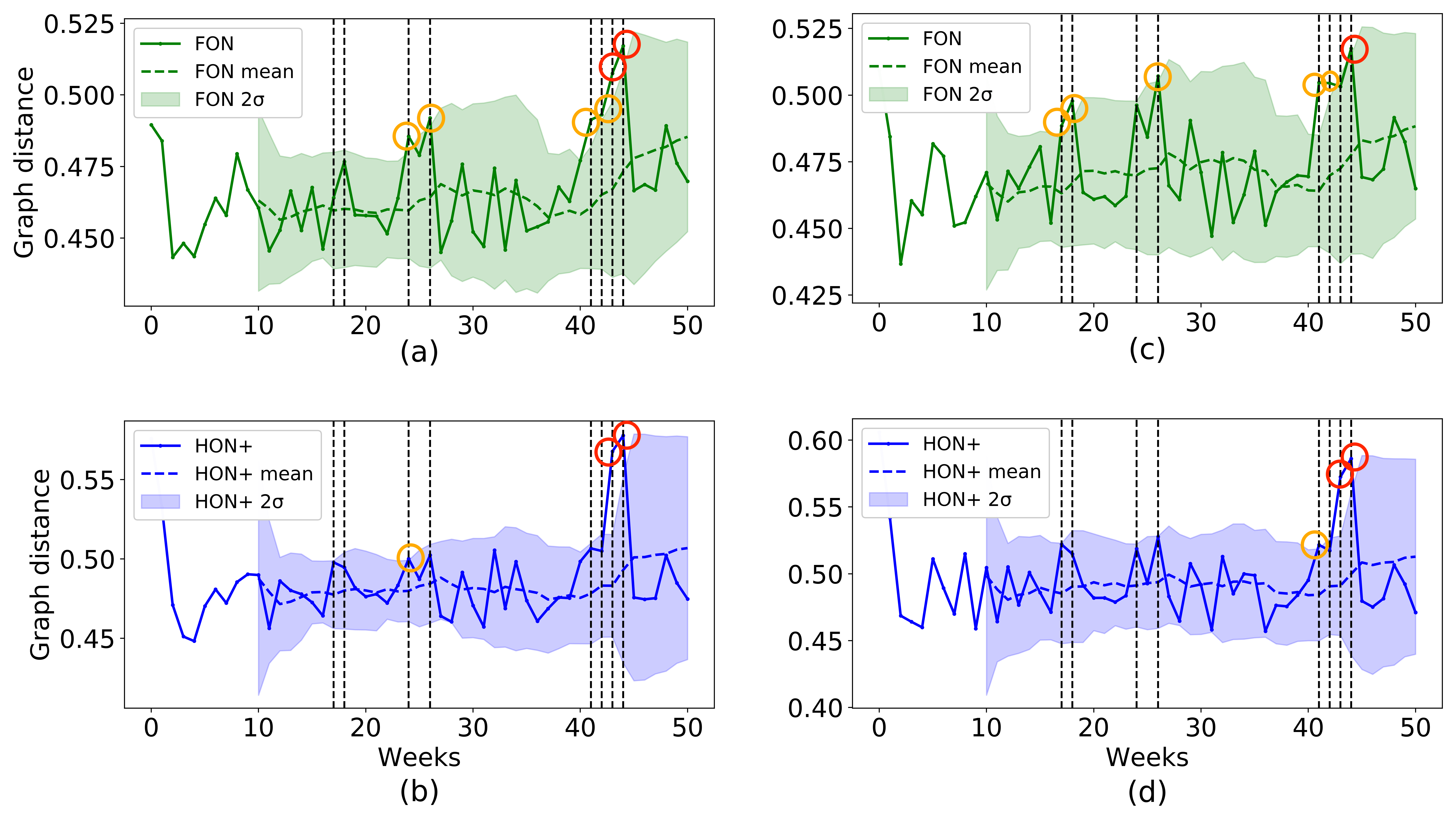}
    \caption{\textbf{Although FON detects the anomalous event (before the addition of the noise) in weeks 43 and 44, it incorrectly shows weeks 24, 26, 41 and 42 as anomalous, resulting in a 4 (out of 6) false positives. HON+, on the other hand, detected the anomalous event with one false positive (week 24) (a). After addition of the noise, FON can no longer detect the true anomaly signal in week 43, and detects the anomaly in week 44 with a graph distance that is very close to the  $2\sigma$ threshold. FON shows false positives in week 17, 18, 26, 41, 42 and one true anomaly, which is also very close to the $2\sigma$ threshold (b).}}
    \label{fig:noise}
\end{figure}

We notice that FON graph distance in week 43 and 44 falls slightly outside the $2\sigma$ threshold as well. However, HON+ deviation from the $2\sigma$ threshold is 3.25 times bigger that of FON in week 44 and 5.2 times bigger in week 43. This becomes important in the presence of noise where anomalies by FON may not be detected. Furthermore, based on FON graph distances, weeks 24, 26, 41, and 42 are also anomalous events (Table~\ref{tab:values} and Fig~\ref{fig:noise}(a)). However, no significant event happened during these weeks. Thus, without any noise, FON can detect anomaly but with a higher false positive rate (4), while HON+ can also correctly detect the anomalies but with only 1 false positive for week 24,  with a very small value above the 2$\sigma$ threshold, as indicated in Table~\ref{tab:values}). 

To illustrate the above point, we designed an experiment to show the robustness of HON+ and FON against noise. We randomly assigned 10\% of all taxis to the next closest police station and constructed the corresponding HON+ and FON. The graph distances before and after adding the noise is shown in  Fig~\ref{fig:noise}(a) and Fig~\ref{fig:noise}(b). The detected anomalies are presented in Table~\ref{tab:values} where the values represent the difference between the graph distance and the $2\sigma$ threshold. We notice that before adding the noise, FON detects the anomalies at week 44 and week 43 with a small margin from the $2\sigma$ threshold. Furthermore, it has a higher false positive rate. 
After adding the noise, FON shows false positives in weeks 17, 18, 26, 41, 42 and one correct anomaly at week 44 which is very close to the 2$\sigma$ threshold (Table~\ref{tab:values}). Furthermore, FON does not detect anything in week 43. HON+, on the other hand, detected the anomalous event (in both weeks 43 and 44) before and after the noise with only one false positive (Table~\ref{tab:values}). 
It is important to note that false positives can be very costly and often require manual correction by human labor.  

{}

\section{Conclusion}

This paper presented a scalable and parameter-free algorithm for extracting higher-order dependencies from the sequential data, and demonstrates the success of higher-order network modeling for anomaly detection in dynamic networks. We show that \HONp is scalable and parameter-free and automates the process of discovering the appropriate variable and higher-order dependencies for each of the nodes in a network.  The complexity analysis of \HONp, as well as running time and memory consumption benchmarking results, demonstrates the scalability of \HONp to large-scale networks.

We further demonstrate that FONs are weak detectors of higher-order anomalies,  especially in the noisy data. 
This emerges because FONs do not adequately capture the sequential orders or indirect pathways in a complex system, thereby providing a limiting view of the behavior of a complex system in their network representation. \HONp can accurately capture such anomalies and also work seamlessly with existing anomaly detection methods to enable more accurate detection of anomalies in comparison to using FON.

\ifx
This paper presents a higher-order network (HON) approach for anomaly detection, which is capable of detecting higher-order anomalies in sequences.
 In this work, we focused on two key elements: efficiency, and accuracy. In terms of accuracy, we show that higher-order anomalies can remain completely undetected using all existing anomaly detection methods that rely on the first-order network (FON). On the efficiency side, we show that despite the advantages of HON, our existing HON construction algorithm is not suitable for anomaly detection due to the extra parameters and poor scalability for big data. We introduce \HONp, a parameter-free algorithm that constructs HON from big data sets.
We show with a large-scale synthetic taxi movement data how multiple existing anomaly detection methods that depend on FON fail to capture higher-order anomalous navigation behaviors; we further show how HON can be plugged into existing methods to enable the detection of various orders of anomalies.
We also demonstrate HON on real-world taxi trajectory data, showing its ability in amplifying higher-order anomaly signals and locating anomalies.
Finally, we provide complexity analysis of \HONp, as well as running time and memory consumption benchmarking results, which demonstrates how one can improve existing anomaly detection approaches using \HONp on large data sets with a small overhead.  
\fi

The higher-order network representation results in a more accurate representation of the underlying trends and patterns in the behavior of a complex system and is the correct way of constructing the network to not miss any important dependencies or signals. This is especially relevant when the data is noisy and has sequential dependencies within indirect pathways. This has numerous applications, ranging from information flow to human interaction activity on a website to transportation to invasive species management to drug and human tracking. 

We note that changes in the HON structure can be more complex than changes in FON, such as the emergence and dissipation of higher-order patterns. In order to use the graph distances that are defined for conventional first-order networks, our current approach treats any changes in the node orders as total removal/addition of that node. This approach may result in more fluctuations in the graph distance and causes HONs to become less overlapping over time. Regardless, measures defined for FONs can still be used for anomaly detection in HONs, since the $2\sigma$ criteria captures the HON fluctuations. One possible improvement can be designing a distance measure for capturing the unique fluctuations in the HON structure. 

Another direction for future work is to classify different types of anomalies given different types of node changes in HON, like the emergence and dissipation of higher-order patterns. In addition to the graph distance metrics, one may also consider structure-based metrics  \cite{akoglu2015graph} that factor in changes of clustering or ranking results and local properties such as motifs on the network. This could be considered as a supervised learning problem, where different categories of anomalies are labeled as classes in the training data and the task is to predict whether those categories of anomalies appear in the testing data. All of these extensions are directly compatible with \HONp, as the resulting HON representation does not impose a change in the network analysis method.

\begin{backmatter}
\section*{List of abbreviations}
HON: Higher-order network\\
FON: First-order network

\section*{Availability of data and materials}
The Portugal taxi data is available through: http://www.geolink.pt/ecmlpkdd2015-challenge/data set.html\\
The Code for generating higher-order network and synthetic data is available at:
https://github.com/msaebi1993/HON-ANOMALY

\section*{Competing interests}
The authors declare that they have no competing interests.

\section*{Funding}
This work is based on research supported in part by the University of Notre Dame Office of Research via Environmental Change Initiative (ECI) and NSF awards EF-1427157 and IIS-1447795; the research was also supported in part by the Army Research Laboratory under Cooperative Agreement no. W911NF-09-2-0053 [the ARL Network Science CTA (Collaborative Technology Alliance)]. 

\section*{Authors' contributions}
All the authors participated in the conception of the research study. JX and MS are equal contributors.  JX and MS implemented the algorithms and performed the experiments. All the authors participated in analysis and writing of the paper. 

\section*{Acknowledgements}
This research was supported in part by NSF Grants IIS-1447795 and CRI-1629914, and by the Army Research Laboratory under Cooperative Agreement no. W911NF-09-2-0053 (the ARL Network Science CTA). The views and conclusions contained in this document are those of the authors and should not be interpreted as representing the official policies, either expressed or implied, of the Army Research Laboratory or the U.S. Government. The U.S. Government is authorized to reproduce and distribute reprints for Government purposes notwithstanding any copyright notation here on.
\end{backmatter}
\bibliographystyle{vancouver.bst}
\bibliography{anomaly}
\section{Figure legends}

\begin{itemize}
    \item Figure 1: Higher-order anomalies cannot be detected by network-based anomaly detection methods if FON is used.
    
    \item Figure 2: Comparison of the {\em active} observation construction in \HON (left) and the {\em lazy} observation construction in \HONp (right, with a much smaller search space). Circled numbers represent the order of execution.
    
    \item Figure 3: Comparing anomaly detection on taxi trajectories based on the first-order dynamic network and the higher-order dynamic network.

    \item Figure 4: \HON is highly sensitive to the size of the data.  For the maximum data size, \HON takes 4.5 times longer than \HONp to run (a), and requires approximately 7.2 times more memory than \HONp (b) . We set  {\em MaxOrder=15} for BuildHON.
    
    \item Figure 5: Given the same data ~\cite{xu2016representing}, \HONp extracts up to $11^{th}$ order in 1/3 run-time and 1/5 memory of \HON. We set {\em MaxOrder=11} for \HON.

    \item Figure 6: Synthetic taxi movement data: variable orders of navigation patterns on 100 cells as a 10x10 grid.
    
    \item Figure 7: Methods that use FON collectively fail to capture anomalous navigation behaviors beyond first-order no matter what distance metric is used, but all show signals when HON is used instead.
    
    \item Figure 8: (a) Labeling of police stations in urban areas of Porto. (b) and (c) the emergence of higher-order traffic patterns in week 43 and 44 (“Burning of the Ribbons” festival) captured by HON, corresponding to the highlighted region in (a).
    
    \item Figure 9: Variation of number of first-order, second-order and third-order nodes in HON of the taxi data of Porto. The anomalous traffic patterns result in a significant change in the number of second and third-order nodes, but not the first-order nodes.
    
    \item Figure 10: The anomalous traffic patterns are more noticeable in HON+ (\tb{c}), while they are not as significant in HON-2 (\tb{b}) and FON (\tb{a}). (\tb{d}) Anomaly detection results on the dynamic network of FON and HON-2. (\tb{e}) Anomaly detection results on the dynamic network of HON-2 and HON+.
    
    \item Figure 11: Although FON detects the anomalous event (before the addition of the noise) in weeks 43 and 44, it incorrectly shows weeks 24, 26, 41 and 42 as anomalous, resulting in a 4 (out of 6) false positives. HON+, on the other hand, detected the anomalous event with one false positive (week 24) (a). After addition of the noise, FON can no longer detect the true anomaly signal in week 43, and detects the anomaly in week 44 with a graph distance that is very close to the  $2\sigma$ threshold. FON shows false positives in week 17, 18, 26, 41, 42 and one true anomaly, which is also very close to the $2\sigma$ threshold (b).
    
    \item Table 1: Represents all the data points in which the graph distance (using FON and HON+) falls outside the $2\sigma$  threshold. The column ``value'' indicates the difference between the graph distance (obtained from FON or HON+) and the closest $2\sigma$  threshold (with respect to the moving average). TP refers to true positive values (correct anomalies), which are marked as bold. FP refers to false positive values. HON+ correctly detects anomalies, results in much lower false positives, and is more robust in presence of noise.
\end{itemize}
\clearpage
\section*{Supplementary materials}
\textbf{Algorithm.} We present the parameter-free and scalable \HONp algorithm for constructing HON. The rule extraction step is given in Algorithm 1, and the network wiring step remains the same as that of HON in~\cite{xu2016representing}. While \HONp algorithm is parameter-free, we provide {\em MaxOrder} and {\em MinSupport } as optional parameters. We also provide the optional {\em ThresholdMultiplier} parameter (the default value 1 is consistent with the HON algorithm), for users to control how aggressive the algorithm prevents higher-order dependencies from being generated. Setting the parameter larger than 1 results in less higher-order dependencies, smaller than 1 for more higher-order dependencies.

\begin{algorithm}[ht]
\begin{spacing}{1}
\small
\caption{\HONp rule extraction algorithm. Given the raw sequential data $T$, extracts arbitrarily high orders of dependencies, and output the dependency rules $R$. Optional parameters include $MaxOrder$, $MinSupport$, and $ThresholdMultiplier$\mbox{}}
\begin{algorithmic}[1]
  \State define \textit{global} $C$ as nested counter
  \State define \textit{global} $D$,$R$ as nested dictionary
  \State define \textit{global} $SourceToExtSource$, $StartingPoints$ as dictionary
  \State
  
  \Function {ExtractRules}{$T$, [$MaxOrder$, $MinSupport$, $ThresholdMultiplier=1$]}
  \State \textit{global} $MaxOrder$, $MinSupport$, $Aggresiveness$
  \State \textsc{BuildFirstOrderObservations}($T$)
  \State \textsc{BuildFirstOrderDistributions}($T$)
  \State \textsc{GenerateAllRules}($MaxOrder$, $T$)

\EndFunction
\State

\Function {BuildFirstOrderObservations}{$T$}
\For {$t$ in $T$}
\For {$(Source, Target)$ in $t$}
\State $C$[$Source$][$Target$] += 1
\State $IC$.add($Source$)
\EndFor
\EndFor
\EndFunction
\State

\Function {BuildFirstOrderDistributions}{$T$}
\For {$Source$ in $C$}
\For {$Target$ in $C[Source]$}
\If {$C[Source][Target] < MinSupport$}
\State $C[Source][Target] = 0$
\EndIf
\For {$Target$ in $C[Source]$}
\If $C[Source][Target] > 0$
\State $D[Source][Target] = C[Source][Target]/(\sum C[Source][*])$
\EndIf
\EndFor
\EndFor
\EndFor

\EndFunction
\State

\Function {GenerateAllRules}{$MaxOrder$, $T$}
\For {$Source$ in $D$}
\State \textsc{AddToRules}($Source$)
\State \textsc{ExtendRule}($Source$, $Source$, 1, $T$)
\EndFor
\EndFunction
\State

\Function {KLDThreshold}{$NewOrder$,$ExtSource$}
\State \Return $ThresholdMultiplier \times NewOrder / log_2(1+\sum C[ExtSource][*])$
\EndFunction

\Function {ExtendRule}{$Valid$, $Curr$, $order$, $T$}
\If {$Order \leq MaxOrder$}
\State \textsc{AddToRules}($Source$)
\Else
\State $Distr=D[Valid]$
\If {$-log_2(min(Distr[*].vals)) < $ \textsc{KLDThreshold}$(order + 1), Curr$}
\State \textsc{AddToRules}($Valid$)
\Else
\State $NewOrder = order + 1$
\State $Extended = $ \textsc{ExtendSource}$(Curr)$
\If {$Extended = \emptyset$}
\State \textsc{AddToRules}($Valid$)
\Else
\For {$ExtSource$ in $Extended$}
\State $ExtDistr = D[ExtSource]$
\State $divergence = $ KLD$(ExtDistr, Distr)$
\If {$divergence > $ \textsc{KLDThreshold}$(NewOrder, ExtSource)$}
\State \textsc{ExtendRule}$(ExtSource,ExtSource,NewOrder,T)$
\Else
\State \textsc{ExtendRule}$(Valid,ExtSource,NewOrder,T)$
\EndIf
\EndFor
\EndIf
\EndIf
\EndIf
\EndFunction

\algstore{honor}
\end{algorithmic}
\end{spacing}
\end{algorithm}

\begin{algorithm}[ht]
\begin{spacing}{1.067}
\caption{\em{(continued)}}
\small
\begin{algorithmic}
\algrestore{honor}

\Function {AddToRules}{Source}:
\For {$order$ in [1..len($Source$) + 1]}
\State $s=Source[0:order]$
\If {not $s$ in $D$ or len($D[s]$) == 0}
\State \textsc{ExtendSource}($s$[1:])
\EndIf
\For {$t$ in $C[s]$}
\If {$C[s][t]>0$}
\State $R[s][t] = C[s][t]$
\EndIf
\EndFor
\EndFor
\EndFunction
\State

\Function {ExtendSource}{$Curr$}
\If {$Curr$ in $SourceToExtSource$}
\State \Return $SourceToExtSource[Curr]$
\Else
\State \textsc{ExtendObservation}$(Curr)$
\If {$Curr$ in $SourceToExtSource$}
\State \Return $SourceToExtsource[Curr]$
\Else
\State \Return $\emptyset$
\EndIf
\EndIf
\EndFunction
\State

\Function {ExtendObservation}{$Source$}
\If {length($Source$) $> 1$}
\If {not $Source[1:]$ in $ExtC$ or $ExtC[Source] = \emptyset$}
\State \textsc{ExtendObservation}$(Source[1:])$
\EndIf
\EndIf
\State $order = length(Source)$
\State define $ExtC$ as nested counter
\For {$Tindex,index$ in $StartingPoints[Source]$}
\If {$index-1 \leq 0$ and $index + order < length(T[Tindex])$}
\State $ExtSource=T[Tindex][index-1:index+order]$
\State $ExtC[ExtSource][Target] += 1$
\State $StartingPoints[ExtSource]$.add(($Tindex, index-1$))
\EndIf
\EndFor

\If {$ExtC=\emptyset$}
\State \Return
\EndIf

\For {$S$ in $ExtC$}
\For {$t$ in $ExtC[s]$}
\If {$ExtC[s][t]<MinSupport$}
\State $ExtC[s][t] = 0$
\EndIf
\State $C[s][t] += ExtC[s][t]$
\EndFor
\State $CsSupport = \sum ExtC[s][*]$
\For {$t$ in $ExtC[s]$}
\If {$ExtC[s][t]>0$}
\State $D[s][t]=ExtC[s][t]/CsSupport$
\State $SourceToExtSource[s[1:]].add(s)$
\EndIf
\EndFor
\EndFor
\EndFunction
\State

\Function {BuildSourceToExtSource}{$order$}
\For {$source$ in $D$}
\If {$len(source) = order$}
\If {$len(source) > 1$}
\State $NewOrder=len(source)$
\For {$starting in [1..len(source)]$}
\State $curr = source[starting:]$
\If {not $curr$ in $SourceToExtSource$}
\State $SourceToExtSource[curr] = \emptyset$
\EndIf
\If {not $NewOrder$ in $SourceToExtSource[curr]$}
\State $SourceToExtSource[curr][NewOrder] = \emptyset$
\EndIf
\State $SourceToExtSource[curr][NewOrder].add(source)$

\EndFor 
\EndIf
\EndIf
\EndFor
\EndFunction


\end{algorithmic}
\label{alg:HON+}
\end{spacing}
\end{algorithm}


\clearpage
\textbf{Effect of {\em MinSupport}.} In this section, we discuss the effect of {\em MinSupport} parameter (in \HON) on the anomaly detection performance. We want to answer the following question: Can we find an optimal value for {\em MinSupport} using parameter sweeping for \HON? Here we show that any values higher than 1 result in lower performance for the anomaly detection using \HON. As a result, all the experiments in the manuscript use {\em MinSupport}=1 for \HON.

We evaluate the anomaly detection performance on the real-world and synthetic data with different values of {\em MinSupport} for \HON. The results for the real-world data are shown in Table 2. We notice that for this data, {\em MinSupport}=1 was the best value, as a higher value led to a decrease in recall and precision both. The total running time for building the HONs (with different values of {\em MinSupport}) was not significant (order of a few seconds with parallel processing), as the size of the networks is small. 

However, the computational time for building the HONs for the synthetic data was significantly higher. The total running time for the parameter sweeping process was 4.372 hours (using parallel processing). On the other hand, the anomaly detection performance was the same for the different {\em MinSupport} values we tried (1 - 5). This is expected since the synthetic data does not have any noisy patterns.

Note that, {\em MinSupport} is a global threshold that applies to all paths regardless of order or support. This threshold should have the ability to vary by path/order/support. For example, we may want to preserve paths that their distribution is very different from their lower-order variant, despite having smaller support. Parameter sweeping for {\em MinSupport} does not necessarily help to preserve significant rules while pruning noise, because it might prune lower orders too aggressively or prune higher-order too mildly. It can also be too aggressive for significantly different paths, or too mild for similar paths. 

\begin{table*}
\small 
\centering
\begin{tabular}{|c|ccccc|}
\hline
{\em MinSupport}  & 1	    & 2	    & 3	        & 4         & 5\\
\hline
Precision	& 0.667	& 0.5	& 0.333	    & 0.167	    & 0.167\\
\hline
Recall	    & 1	&   0.5	    & 0.5	    & 0.5	    & 0.5\\
\hline
    \end{tabular}
    \caption{Precision and recall values for the anomaly detection task using BuildHON with different values of {\em MinSupport}.}
    \label{tab:min_supp}
\end{table*}
\end{document}